\documentclass[twocolumn,prl,showpacs,preprintnumbers,amsmath,amssymb]{revtex4}

\usepackage{graphicx}% Include figure files
\usepackage{dcolumn}% Align table columns on decimal point
\usepackage{bm}% bold math
\usepackage{subfigure}
\usepackage{float}

\begin{document}
%\preprint{APS/123-QED}
\title{\textbf{\textrm{Spike-timing-dependent plasticity with axonal delay tunes networks of Izhikevich neurons to the edge of synchronization transition with scale-free avalanches}}}
\author{Mahsa Khoshkhou}
\author{Afshin Montakhab}
\email{montakhab@shirazu.ac.ir}
\affiliation{Department of Physics, College of Sciences, Shiraz University, Shiraz 71946-84795, Iran}
\date{\today}
%%%-----------------------------------------------------------------------------
\begin{abstract}
Critical brain hypothesis has been intensively studied both in
experimental and theoretical neuroscience over the past two
decades. However, some important questions still remain: (i) What
is the critical point the brain operates at? (ii) What is the
regulatory mechanism that brings about and maintains such a
critical state? (iii) The critical state is characterized by
scale-invariant behavior which is seemingly at odds with
definitive brain oscillations?  In this work we consider a
biologically motivated model of Izhikevich neuronal network with
chemical synapses interacting via spike-timing-dependent
plasticity (STDP) as well as axonal time delay. Under generic and
physiologically relevant conditions we show that the system is
organized and maintained around a synchronization transition point
as opposed to an activity transition point associated with an
absorbing state phase transition. However, such a state exhibits
experimentally relevant signs of critical dynamics including
scale-free avalanches with finite-size scaling as well as
branching ratios. While the system displays stochastic
oscillations with highly correlated fluctuations, it also displays
dominant frequency modes seen as sharp peaks in the power
spectrum.  The role of STDP as well as time delay is crucial in
achieving and maintaining such critical dynamics, while the role
of inhibition is not as crucial. In this way we provide definitive
answers to all three questions posed above. We also show that one
can achieve supercritical or subcritical dynamics if one changes
the average time delay associated with axonal conduction.

\end{abstract}
%%%-----------------------------------------------------------------------------

%%%%%PACS means Physics and Astronomy Classification Scheme
%\keywords{}%Use showkeys class option if keyword
\maketitle
%--------------------------------------------------------------------------------

\section{Introduction}

Since its inception nearly two decades ago, the critical brain
hypothesis has gained a considerable amount of attention in the
literature \cite{Plenz2014,Legenstein2008,Chialvo2010}. Although
it has encountered some skepticism at times \cite{Beggs2012}, it
has now grown to a relatively mature field with substantial body
of theoretical and experimental evidence to support it
\cite{Beggs2003,Beggs2004,Plenz2007,Beggs2012,Chialvo2013,Friedman2012,PRL2019,Levina2007}.
Brain criticality is thought to underlie many of its fundamental
properties such as optimal response, learning, information
storage, as well as transfer \cite{Hesse2014}.  The original ideas
of brain criticality came out of studies of self-organized
criticality, where a threshold dynamics leads to a balance between
slow drive and fast dissipation in open nonequilibrium systems and
thus observation of critical dynamics \cite{BTW}. It is now
generally believed that long-term evolution has led to a balance
between excitatory as well as inhibitory tendencies which place
the brain ``on the edge", i.e. a critical point. However, this
does not necessarily answer the problem of stability of the
critical state, as some neurophysiological mechanism is needed to
maintain the system near the critical point against many possible
perturbative effects.

It seems like there are some important theoretical issues which
have remained open in regards to brain criticality: (i) What
exactly is the phase transition which determines the critical
point? Traditionally, this has been assumed to be the
absorbing-state phase transition motivated by the studies of
self-organized criticality \cite{Mont98,VDMZ2000}. However, in
some recent studies, it has been indicated that the brain is
maintained near a synchronization transition \cite{PRL2019,di
Santo2018}. We note that some authors have also argued for the
existence of the extended critical region similar to that of
``Griffiths phase" \cite{Moretti2013,Meunoz2010,Odor2015,MMV2017}.
However, such critical regions also typically occur near the
absorbing phase transition where the system transitions from an
inactive phase to an active phase. (ii) What is the
self-organizing mechanism which leads to, and maintains the system
in a critical state? As mentioned above the balance between
excitatory and inhibitory tendencies are thought to be the long
time solution to this question. However, physiological mechanism
such as synaptic plasticity are also considered to be important
mechanism to maintain the nervous system in a balanced state on
shorter time scales. Clearly, extended criticality can also
alleviate such a problem to a certain extend as criticality is
observed for a range of parameter instead of a particular point.
(iii) If the brain is in the critical state with its associated
scale-invariant behavior, how could it also display definitive
rhythmic behavior via brain oscillations?

Brain plasticity is increasingly being recognized as an important
and fundamental property of a healthy nervous system.  In
particular, spike-timing-dependent-plasticity (STDP) is an
important mechanism which can modify synaptic weights on very
short time scales. Therefore, it seems reasonable to invoke STDP
as a self-organizing mechanism. In a STDP protocol, the strength
of a synapse is modified based on the relative spike-timing of its
corresponding pre- and post-synaptic neurons, i.e., STDP
incorporates the causality of pre- and post-synaptic spikes into
the synaptic strength modifications. If the pre-synaptic neuron
spikes first and leads to the post-synaptic neuron to spike
shortly afterward, then the synapse is potentiated. Reversely, if
the pre-synaptic spike follows the post-synaptic spike the synapse
will be depressed \cite{Markram2012,Song2000,Bi2001,Sjostrom2010}.
The competition between coupling and decoupling forces arising
from successive potentiation and depression of synapses tunes the
neural network into a balanced dynamical state.

Our work in this paper is motivated by the above considerations.
In particular, we propose to study a biologically plausible model
of cortical networks, i.e. Izhikevich neurons, along with
neurophysiological regulatory mechanism such as STDP with suitable
axonal conduction delays in order to answer some of the above
posed questions. Interestingly, we find that our regulatory system
self-organizes the neuronal network to the ``edge of
synchronization" in physiologically meaningful parameter regime.
We first establish some of the characteristics of such a steady
state. More importantly, we look for characteristics of critical
dynamics in such a minimally synchronized steady state. Motivated
by various experiments, we look for neuronal avalanches, branching
ratios, as well as power spectrum of activity time-series.  We
find that such a system on the edge of synchronization exhibits
significant indications of critical dynamics including
scale-invariant avalanches with finite-size scaling. Our results
provide definitive answers to the above questions in a
biologically plausible model of neuronal networks.

In the following section, we describe the model we use for our
study. Results of our numerical study is presented in section III,
and we close the paper with some concluding remarks in section IV.

\section{Model and Methods}
The studied cortical networks consist of $N$ spiking Izhikevich
neurons which interact by transition of chemical synaptic currents
with axonal conduction delays. The dynamics of each neuron is
described by a set of two differential equations
\cite{Izhikevich2003}:
\begin{equation}\label{sequ1}
\frac{dv_i}{dt}=0.04v_i^2+5v_i+140-u_i+I_i^{DC}+I_i^{syn}
\end{equation}
\begin{equation}\label{sequ2}
\frac{du_i}{dt}=a(bv_i-u_i)
\end{equation}
\\with the auxiliary after-spike reset:
\begin{equation}\label{sequ3}
\text{if}\ \ v_i{\geq}30,\ \ \text{then} \ v_i \ {\rightarrow} \ c \ \ \text{and}\ u_i \ {\rightarrow} \ u_i+d
\end{equation}
\\
for $i=1, 2,..., N$. Here $v_i$ is the membrane potential and
$u_i$ is the membrane recovery variable. When $v_i$ reaches its
apex ($v_{max}=30$ mV), voltage and recovery variable are reset
according to Eq.(4). $a$, $b$, $c$ and $d$ are four adjustable
parameters in this model. Tuning these parameters, Izhikevich
neuron is capable of reproducing different intrinsic firing
patterns observed in real excitatory and inhibitory neurons
\cite{Izhikevich2003}. We set these parameters so that excitatory
neurons spike regularly and inhibitory neurons produce fast
spiking pattern
\cite{Izhikevich2007,Izhikevich2003,Izhikevich2006}.

\begin{table*}[!htbp]
\caption{\small Values of constant parameters used in this study.}
\label{table1}
\begin{center}
\begin{tabular}{||m{3.0cm}|m{1.7cm} m{1.7cm} m{1.7cm} m{1.7cm} m{1.7cm} m{1.7cm} m{1.7cm} m{1.5cm}||}
\hline \hline
Izhikevich neuron & $a_{ex}=0.02$ & $b_{ex}=0.2$ & $c_{ex}=-65$ & $d_{ex}=8$& $a_{in}=0.1$ & $b_{in}=0.2$ & $c_{in}=-65$ & $d_{in}=2$ \\
\hline
Synaptic current & $\tau_f=0.2$ &  $\tau_s=1.7$ & $V_{0,ex}=0$ & $V_{0,in}=-75$ & & & &  \\
\hline
STDP rule & $A_+=0.05$ & $A_-=0.05$ &  $\tau_+=20$ & $\tau_-=20$ & $g_{min}=0$ & $g_{max}=0.6$ &  & \\
\hline \hline
\end{tabular}
\end{center}
\end{table*}

The term $I_i^{DC}$ is an external current which determines
intrinsic firing rate of uncoupled neurons. Regularly spiking
Izhikevich neurons exhibits a Hopf bifurcation at $I^{DC}=3.78$
\cite{KM2018}. We choose values of $I_i^{DC}$ randomly from a
Poisson distribution with the mean value $10$.  The term
$I_i^{syn}$ represents the chemical synaptic current delivered to
each post-synaptic neuron $i$ \cite{Roth}:
\begin{equation}\label{sequ4}
I_i^{syn}=\frac{V_0-v_i}{D_i}\sum_jg_{ji}\frac{exp(-\frac{t-(t_j+\tau_{ji})}{\tau_s})-exp(-\frac{t-(t_j+\tau_{ji})}{\tau_f})}{\tau_s-\tau_f}
\end{equation}
\\
Here $D_i$ is the in-degree of node $i$, $t_j$ is the instance of
last spike of pre-synaptic neuron $j$, and $\tau_{ji}$ is the
axonal conduction delay from pre-synaptic neuron $j$ to
post-synaptic neuron $i$. If axonal delays are not taken into
account, then $\tau_{ji}=0$ for all $j\neq i$. Axonal delay values
of $\tau_{ji}$ are chosen randomly from a Poisson distribution
with mean value $\tau=\langle \tau_{ji} \rangle$.  $\tau_f$ and
$\tau_s$ are the synaptic fast and slow time constants and $V_0$
is the reversal potential of the synapse. If inhibition is
included, then motivated by the properties of cortical networks
\cite{DeFelipe1993}, we set population density of inhibitory
neurons to twenty percent, i.e. $\alpha=0.2$ while the initial
strength of inhibitory synapses are chosen four times the strength
of excitatory synapses. Therefore, the excitation-inhibition ratio
is balanced.  $\alpha=0$ indicates that we are only considering a
network of excitatory neurons. $g_{ji}$ is the corresponding
element of the adjacency matrix of the network which denotes the
strength of synapse from pre-synaptic neuron $j$ to post-synaptic
neuron $i$. Each type of synapses are initially static and have
equal strength. $g_{ji}=g_s$ if neurons $j$ and $i$ are connected
and the synapse is excitatory, $g_{ji}=4g_s$ if neurons $j$ and
$i$ are connected and the synapse is inhibitory, and $g_{ji}=0$
otherwise. When we turn the STDP on, strength of \emph{excitatory}
synapses are modified according to a soft-bound STDP rule
\cite{Markram2012,Song2000,Bi2001,Sjostrom2010}, while the
strength of inhibitory synapses are fixed. If pre-synaptic neuron
$j$ fires a spike at time $t=t_{pre}$, then the strength of
synapse is modified to $g_{ji} \rightarrow g_{ji}+\Delta g_{ji}$,
where:
\begin{eqnarray}\label{sequ5}
\Delta g_{ji}= \left \{
\begin{array}{cc}
A_+(g_{max}-g_{ji})e^{-\frac{\Delta t-\tau_{ji}}{\tau_+}} & \text{if} \ \Delta t>\tau_{ji}  \\
-A_-(g_{ji}-g_{min})e^{\frac{\Delta t-\tau_{ji}}{\tau_-}} & \text{if} \ \Delta t \leq \tau_{ji}
\end{array}
\right.
\end{eqnarray}
\\
Here, $\Delta t=t_{post}-t_{pre}$ is the time difference of last
post- and pre-synaptic spikes, $A_+$ and $A_-$ determine the
maximum synaptic potentiation and depression, $\tau_+$ and
$\tau_-$ determine the temporal extent of the STDP window for
potentiation and depression, and  $g_{min}$ and $g_{max}$ are the
lower and upper bounds of synaptic strength. The values of all
parameters for Izhikevich neuron, synaptic current and STDP rule
are listed in Table I.

\begin{figure}[!htbp]
\begin{center}
{\includegraphics[width=0.45\textwidth,height=0.25\textwidth]{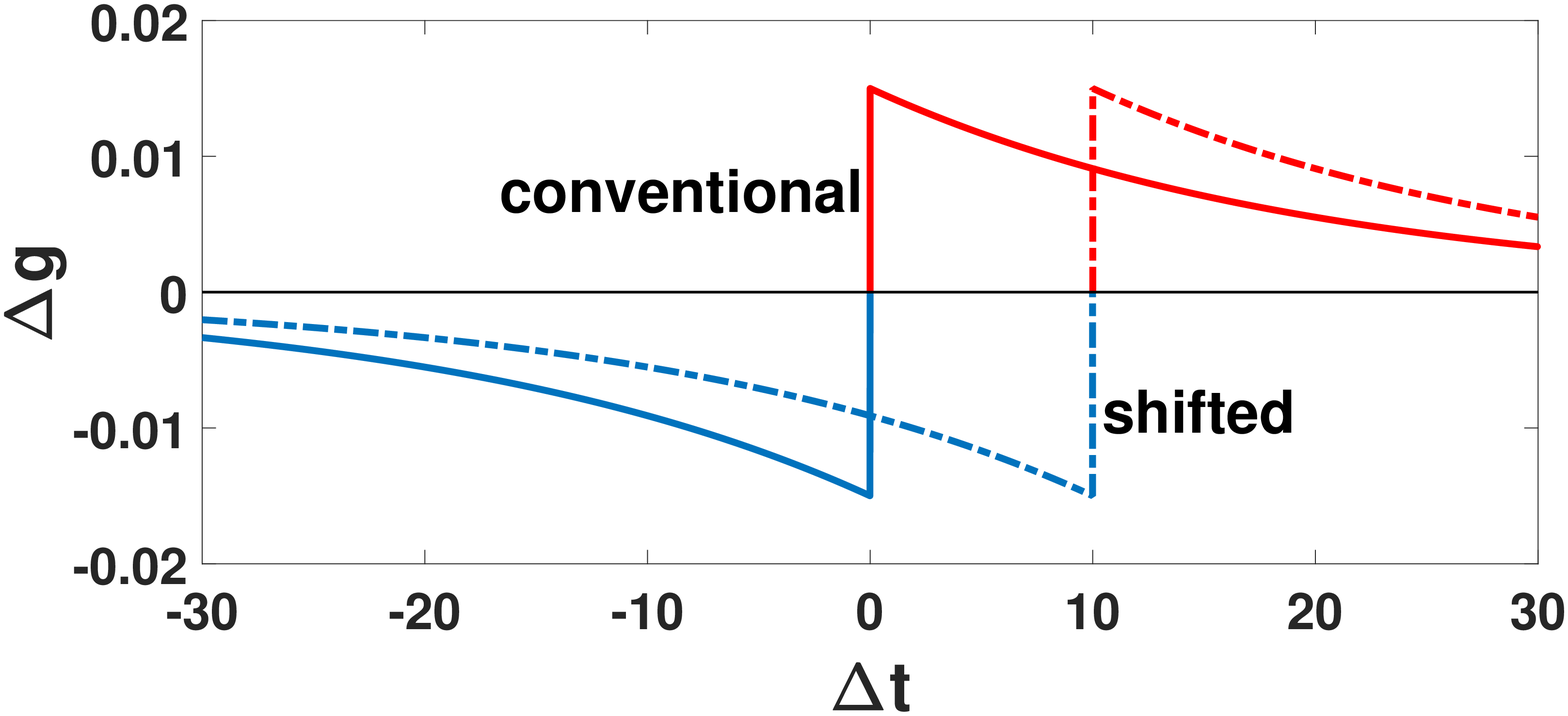}}
\end{center}
\caption{\small Conventional (solid line) and shifted (dashed line) STDP temporal window function $\Delta g= f(\Delta t)$. Blue parts denote depression and red parts denote potentiation. Units of $\Delta t$ is ms.} \label{fig1}
\end{figure}

We consider a temporally shifted STDP window for which the
boundary separating potentiation and depression does not occur for
simultaneous pre- and post-synaptic spikes, but rather for spikes
separated by a small time interval \cite{Babadi2010}. We set the
value of this shift equal with the actual axonal delay for each
synapse. This rule retrieves the conventional STDP rule when no
time-delay is considered, $\tau_{ji}=0$. We have plotted the STDP
temporal window function $\Delta g= f(\Delta t)$ and its shift in
Fig.\ref{fig1}. This temporal shift causes synchronous or
nearly-synchronous pre- and post-synaptic spikes to induce
long-term depression.

We integrate the dynamical equations using fourth-order
Runge-Kutta method with a time step $h=0.01\text{ms}$ and obtain
$v_i(t)$. We typically evolve the entire system for a long time
and make sure that the system has reached a stationary state. We
then perform our measurements and calculations. We obtain the
instants of firings of all neurons and then assign a phase to each
neuron between each pairs of successive spikes
\cite{Pikovsky1997}:
\begin{equation}\label{sequ6}
\phi_i(t)=2\pi\frac{t-t_i^m}{t_i^{m+1}-t_i^m}
\end{equation}
\\
while $t_i^m$ is the time that neuron $i$ emits its $m^{th}$ spike. We define a time-dependent order parameter:
\begin{equation}\label{sequ7}
S(t)=\frac{2}{N(N-1)}\sum_{i\neq j}cos^2\Big{(}\frac{\phi_i(t)-\phi_j(t)}{2}\Big{)}
\end{equation}
\\
This order parameter measures the collective phase synchronization
at time $t$. $S(t)$ is bounded between 0.5 and 1. If neurons spike
out-of-phase, then $S(t){\simeq}0.5$, when they spike completely
in-phase $S(t){\simeq}1$ and for states with partial synchrony
$0.5<S(t)<1$. The global order parameter $S^*$ is the
long-time-average of $S(t)$ at the stationary state after the
influence of STDP ($S^*=\langle S(t)\rangle_t$). We note that the
intricate details of the model along with the need to obtain
long-time dynamics of the system, limit our computational
abilities. We have therefore performed simulations for
$100<N<1000$.  We find that our general results and conclusions
are independent of the system size used and therefore report most
of our results for $N\approx 500$. In the next section we will
present a systematic study of the system above, paying particular
attention to the effect of STDP, time delay, and inhibition.

\section{Results}
Spiking Izhikevich neurons with static chemical synapses exhibit a
continuous transition to phase synchronization upon increasing
synaptic strength, i.e. the amount of global synchrony depends on
the average synaptic strength \cite{KM2018}. Now, consider the
simple case of an all-to-all network of excitatory neurons without
axonal delays. STDP is off initially. $S(t)$ timeseries for
different values of $g_s$ are illustrated in Fig.2(a). It is
observed that $S(t)$ depends on $g_s$ as is expected. Next, we
turn on the STDP at $t=5\text{s}$. Interestingly, it is seen that
$S(t)$ timeseries evolve to a common state regardless of their
initial values. Thus, as STDP modifies the synaptic strengths,
neural network organizes into a final state with a specific global
phase synchronization $S^*$ independent of the initial synaptic
strengths. Our investigations reveal that this is a generic
condition emerging in neural networks with different underlying
structures. We also find that the amount of $S^*$ is independent
of many parameters including the amplitudes and time extents of
STDP rule, and intrinsic firing rate of neurons. However $S^*$
depends drastically on the average value of axonal conduction
delays. Fig.2(b) shows that increasing $\tau$ leads to a phase
transition from strongly synchronized states with $S^*\simeq 1$ to
asynchronous states with $S^*\simeq 0.5$, for neural networks with
$\alpha=0$ and $\alpha=0.2$. Fig.2(b) also shows that inhibition
has a secondary role in the amount of steady state synchronization
,$S^*$, as compared to axonal delay, $\tau$. Important to our
purposes, it shows that for $\tau=10\text{ms}$ the systems stand
at the boundary of phase synchronization for both $\alpha$
values.Note the importance of time delay as it causes STDP to
depress (weaken) the synchronous neurons, thus reducing the amount
of $S^*$ in the system.

\begin{figure}[!htbp]
\begin{center}
\subfigure{\includegraphics[width=0.45\textwidth,height=0.20\textwidth]{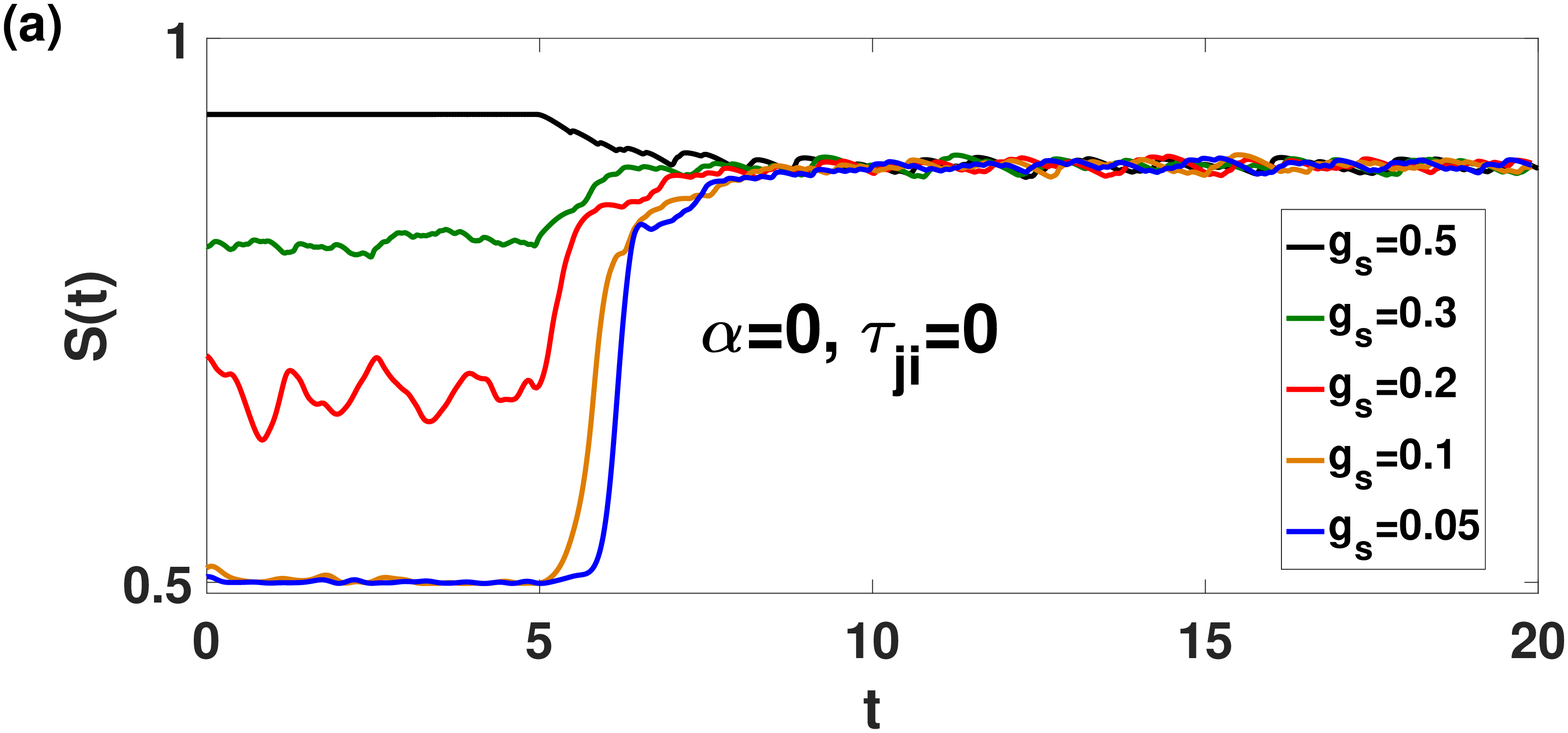}\label{fig2a}}
\subfigure{\includegraphics[width=0.45\textwidth,height=0.20\textwidth]{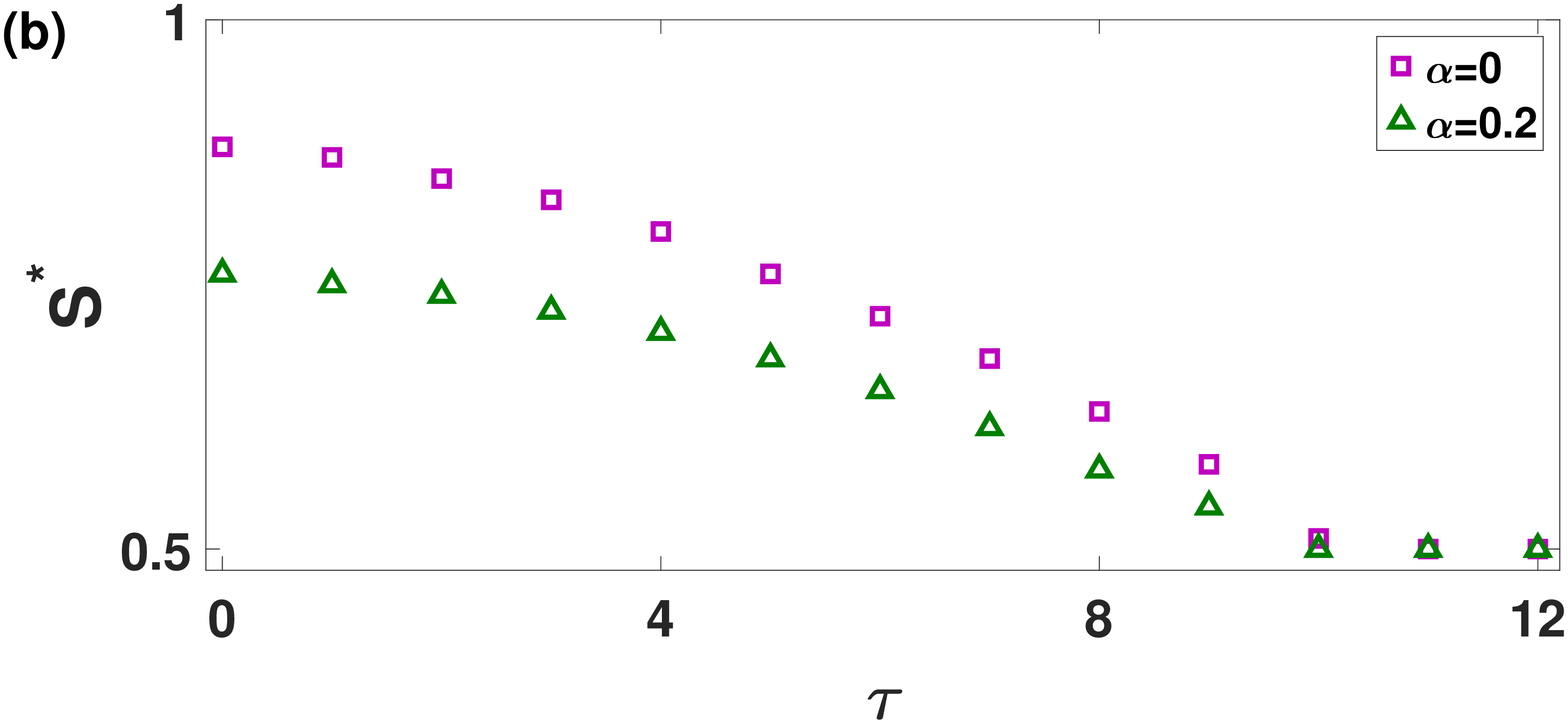}\label{fig2b}}
\end{center}
\caption{\small (a) Effect of STDP on the time evolution of $S(t)$
for all-to-all networks of excitatory spiking neurons with
different $g_s$. The unit of time axis is in seconds. (b)
Dependence of $S^*$ on $\tau$ for $\alpha=0$ and $\alpha=0.2$.}
\label{fig2}
\end{figure}

In order to further investigate the properties of Izhikevich
neuronal networks, we consider four different networks of $N=500$:
(1) a network of purely excitatory neurons without time-delay
($\alpha=0$, $\tau_{ji}=0$), (2) a network of purely excitatory
neurons with axonal conduction delays ($\alpha=0$,
$\tau=10\text{ms}$), (3) a network of excitatory and inhibitory
neurons without time-delay ($\alpha=0.2$, $\tau_{ji}=0$), and (4)
a network of excitatory and inhibitory neurons with axonal
conduction delays ($\alpha=0.2$, $\tau=10\text{ms}$). We have
studied networks with different $\tau$ values, but we display
mostly the results in cases for which all delays are zero
($\tau_{ji}=0$) and $S^* \gg 0.5$ as well as those with $\tau=10
\text{ms}$ for which $S^* \rightarrow 0.5^+$. We note that while
our results (Fig.2) show that $\tau=10 \text{ms}$ is an
interesting case of transition point, such an actual value for
axonal delay is experimentally meaningful \cite{Swadlow2012}. We
turn on STDP at $t=5\text{s}$ in a complete network and monitor
its influence on different features of each system.

\subsection{Synchronization and average synaptic weights}
\begin{figure}[!htbp]
\begin{center}
\subfigure{\includegraphics[width=0.22\textwidth,height=0.15\textwidth]{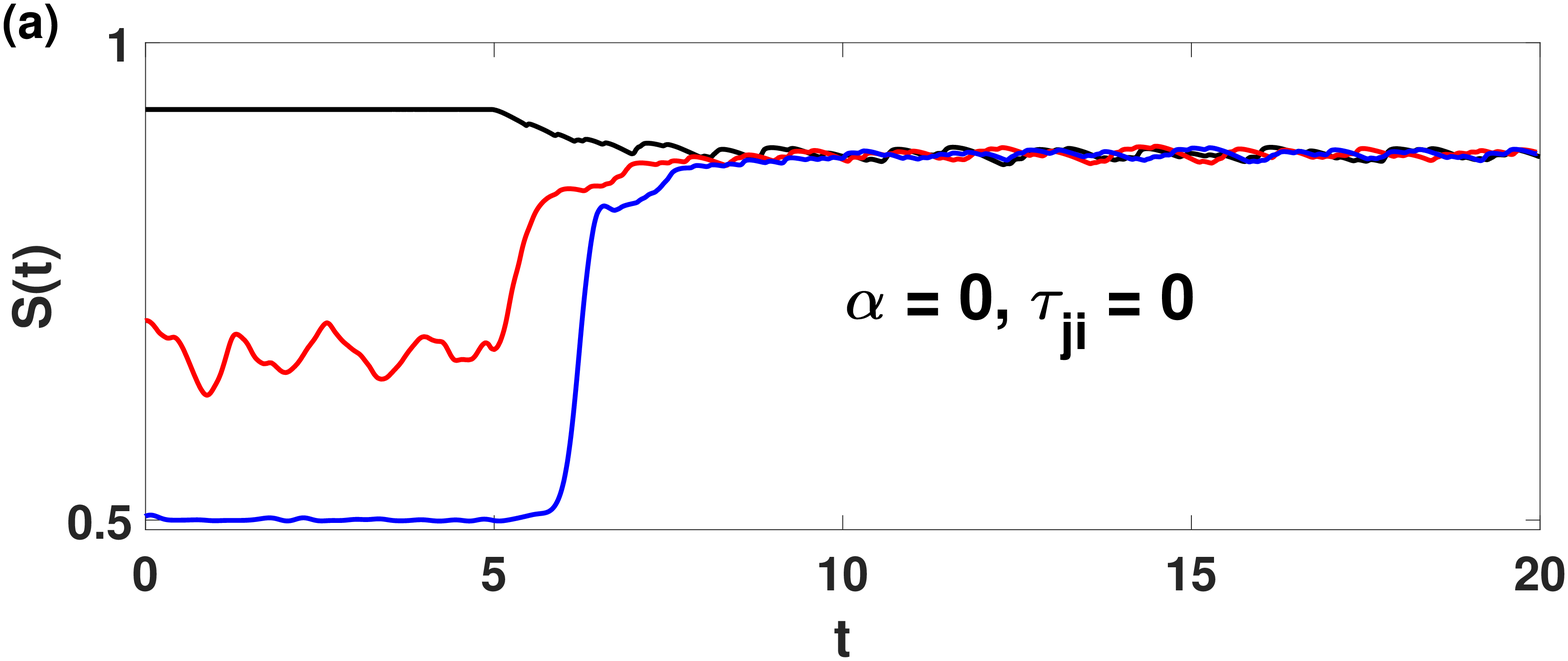}\label{fig3a}}
\subfigure{\includegraphics[width=0.22\textwidth,height=0.15\textwidth]{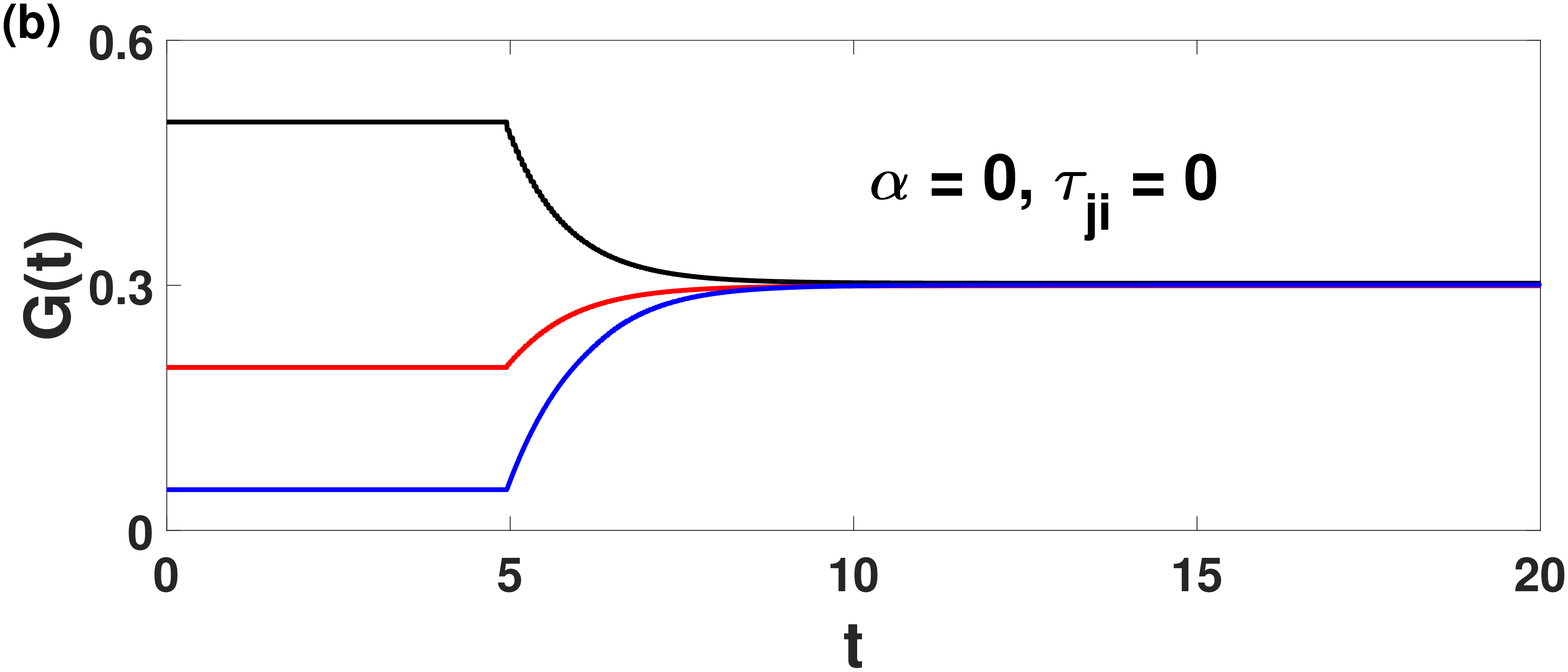}\label{fig3b}}
\subfigure{\includegraphics[width=0.22\textwidth,height=0.15\textwidth]{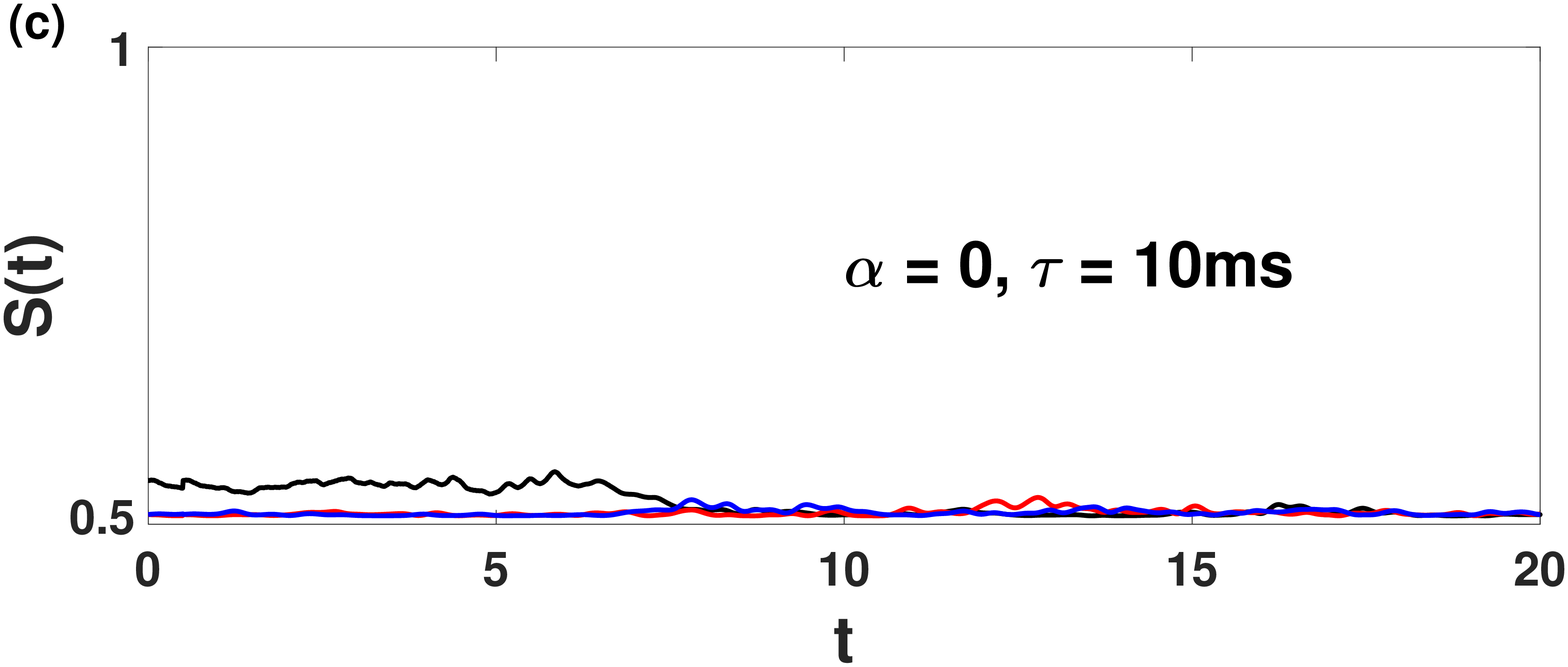}\label{fig3c}}
\subfigure{\includegraphics[width=0.22\textwidth,height=0.15\textwidth]{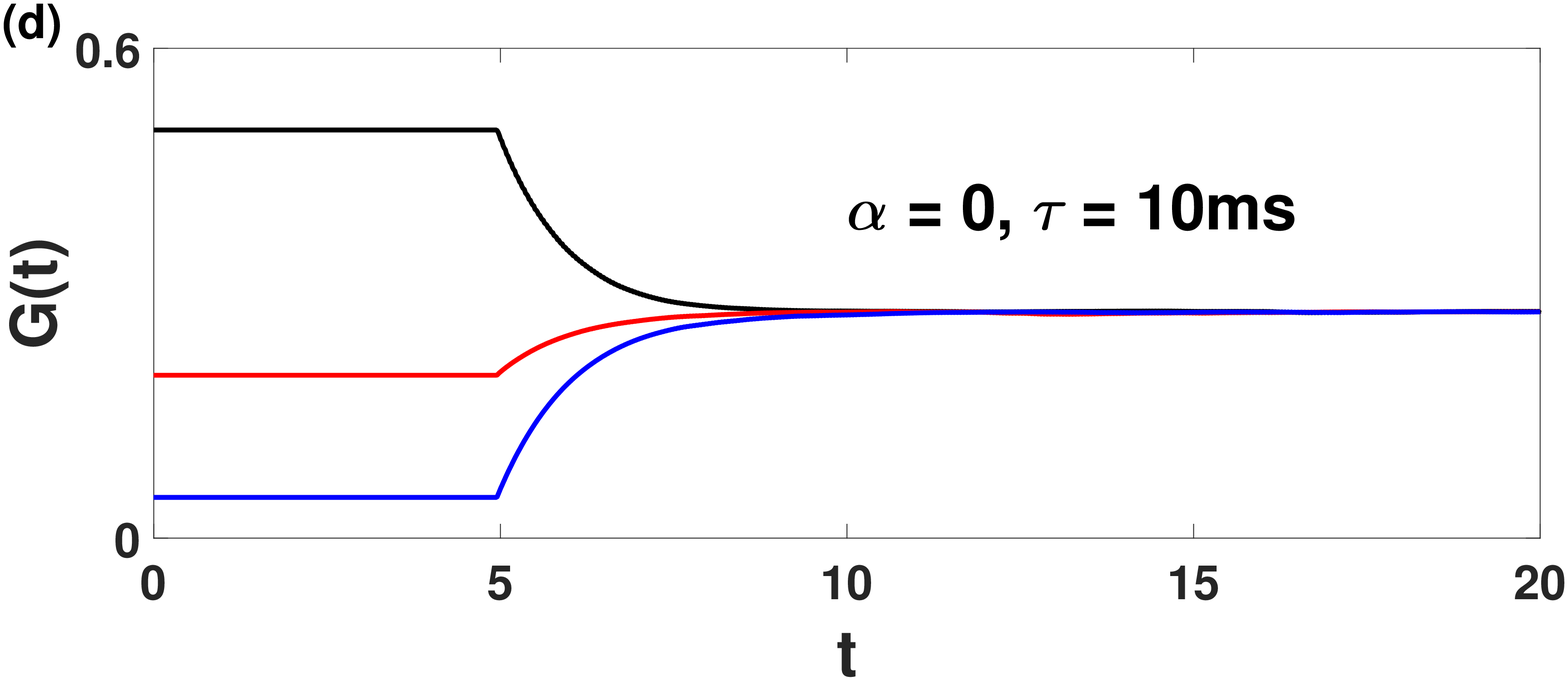}\label{fig3d}}
\subfigure{\includegraphics[width=0.22\textwidth,height=0.15\textwidth]{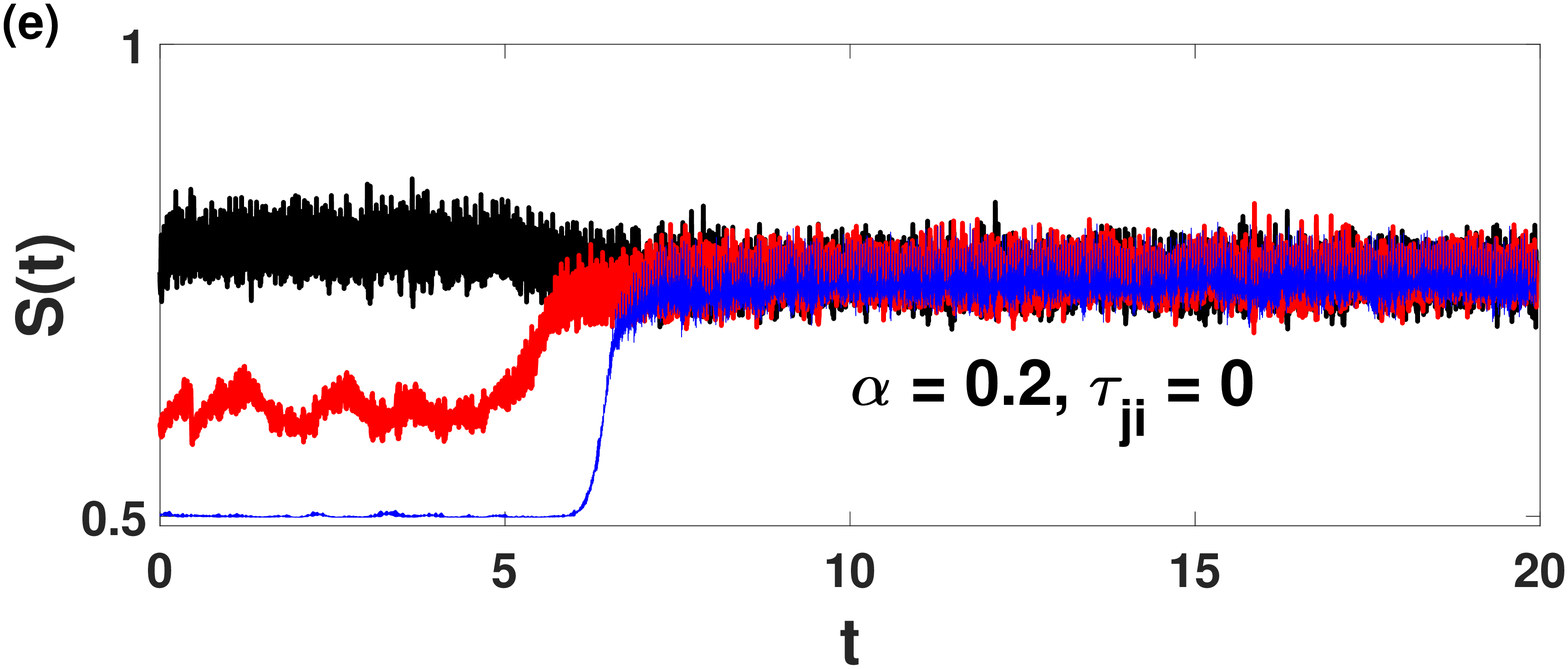}\label{fig3e}}
\subfigure{\includegraphics[width=0.22\textwidth,height=0.15\textwidth]{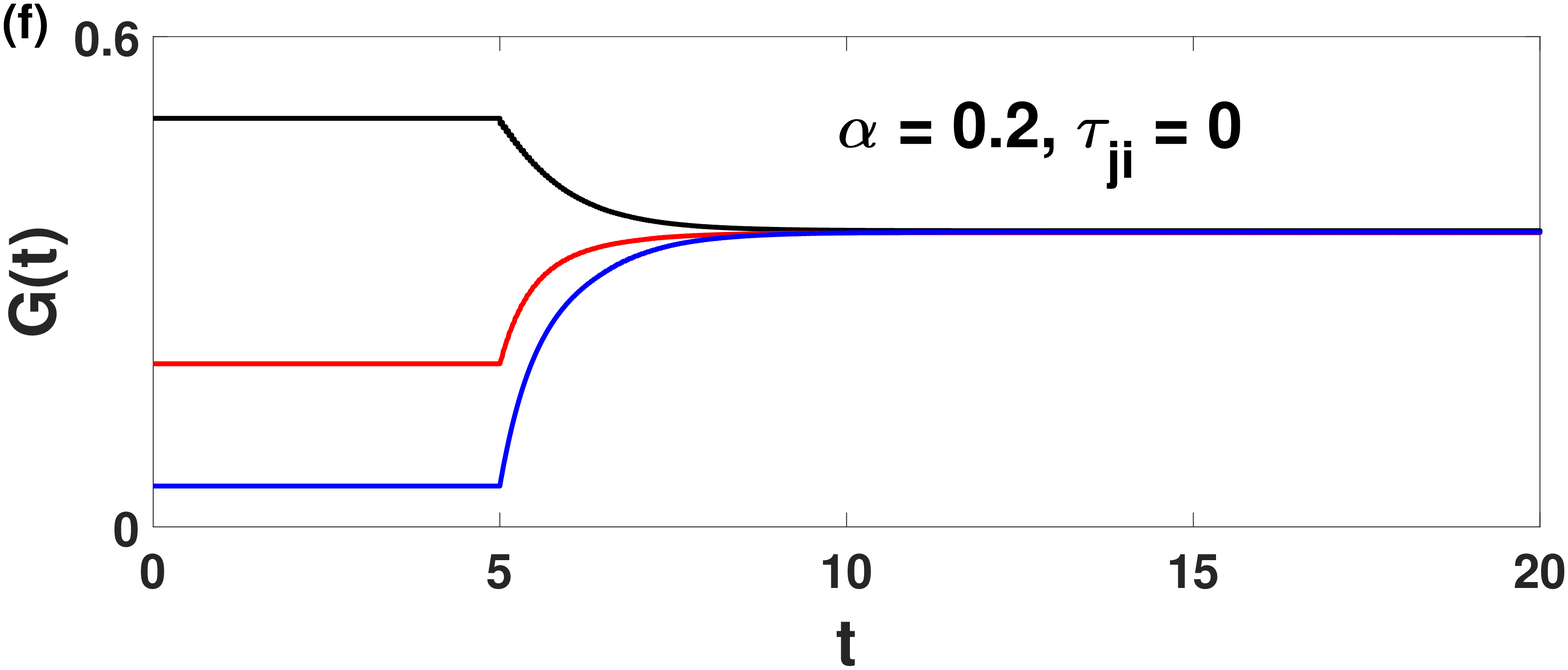}\label{fig3f}}
\subfigure{\includegraphics[width=0.22\textwidth,height=0.15\textwidth]{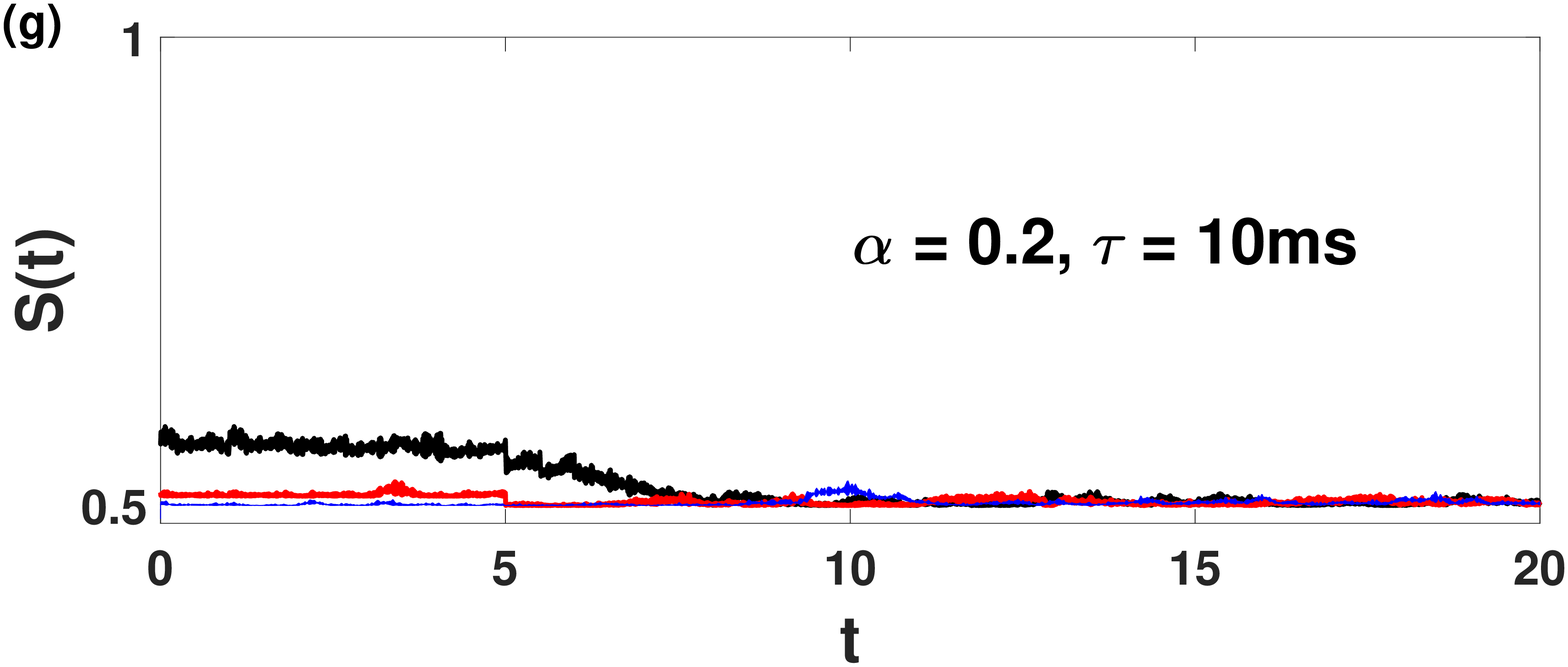}\label{fig3g}}
\subfigure{\includegraphics[width=0.22\textwidth,height=0.15\textwidth]{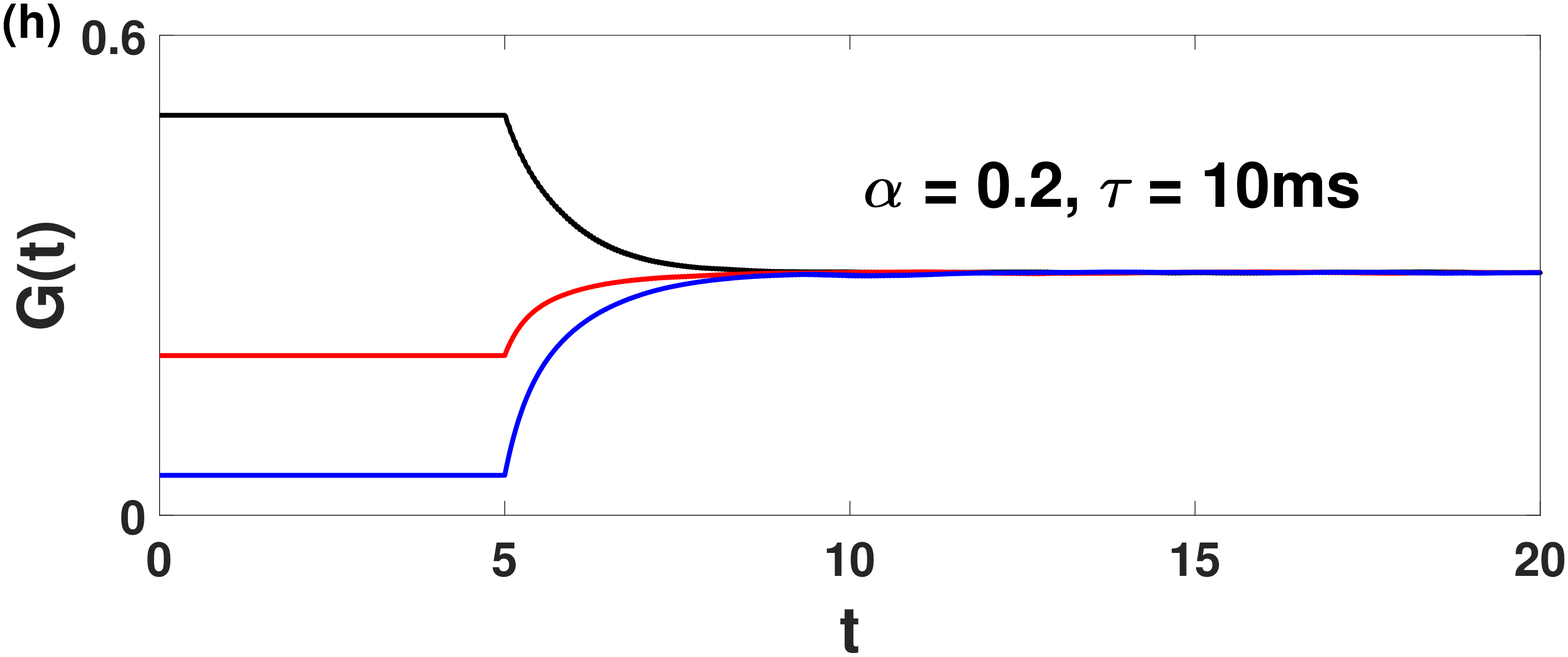}\label{fig3h}}
\end{center}
\caption{\small Timeseries of $S(t)$ and $G(t)$ and the influence
of STDP on them. The unit of time axis is in seconds. STDP is
turned on at $t=5\text{s}$. In each panel different line colors
demonstrate different static synaptic strengths, $g_s=0.5$
(black), $g_s=0.2$ (red) and  $g_s=0.05$ (blue).} \label{fig3}
\end{figure}

The influence of STDP on the timeseries $S(t)$ in different
conditions is illustrated in the left column of Fig.3. Each panel
contains three plots with different values of $g_s$, i.e. the
initial synaptic weights. When STDP is off, $S(t)$ depends on
$g_s$. Turning STDP on, each system reaches a final state with a
specific amount of synchronization $S^*$, regardless of initial
level of order (regardless of $g_s$). However, $S^*$ depends on
$\tau$ and $\alpha$. Systems (1) and (3) reach a strongly
synchronized states with $S^*\simeq0.88$ and $S^*\simeq0.75$,
respectively. Implementation of conduction delays drive the
dynamics toward lower levels of order. Systems (2) and (4) with
$\tau=10\text{ms}$ lead to states at the edge of transition with
$S^*\simeq0.509$ and $S^*\simeq0.503$, respectively. The right
column of Fig.3 represents the timeseries of the average strength
of excitatory synapses, for the corresponding system in the left
column represented by, $G(t)=\frac{1}{N_L}\sum_{j\neq
i}g_{ji,ex}(t)$, where $N_L$ is the number of existing excitatory
links. It is observed that at the final states $G(t)\simeq0.3$ for
all the systems. It is interesting that the final average value of
synaptic weight is independent of the amount of inhibition and
and/or axonal delay, as well as initial distribution. However, the
main point here is that the amount of synchronization in the
system is not solely determined by average synaptic strength but
crucially depends on axonal conduction delay, and to a lesser
degree on inhibition.

\subsection{Synaptic distributions}
\begin{figure}[!htbp]
\begin{center}
\subfigure{\includegraphics[width=0.22\textwidth,height=0.15\textwidth]{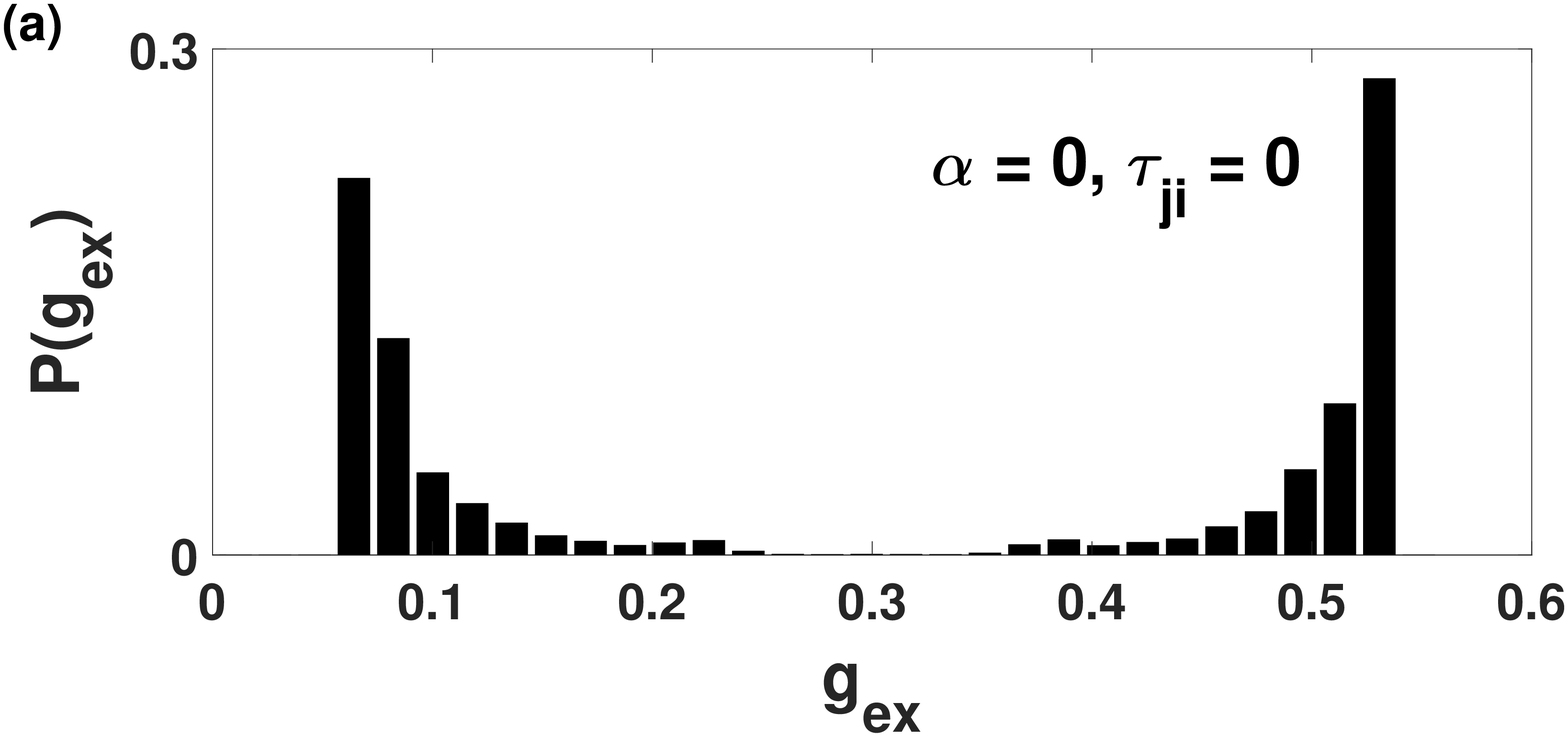}\label{fig4a}}
\subfigure{\includegraphics[width=0.22\textwidth,height=0.15\textwidth]{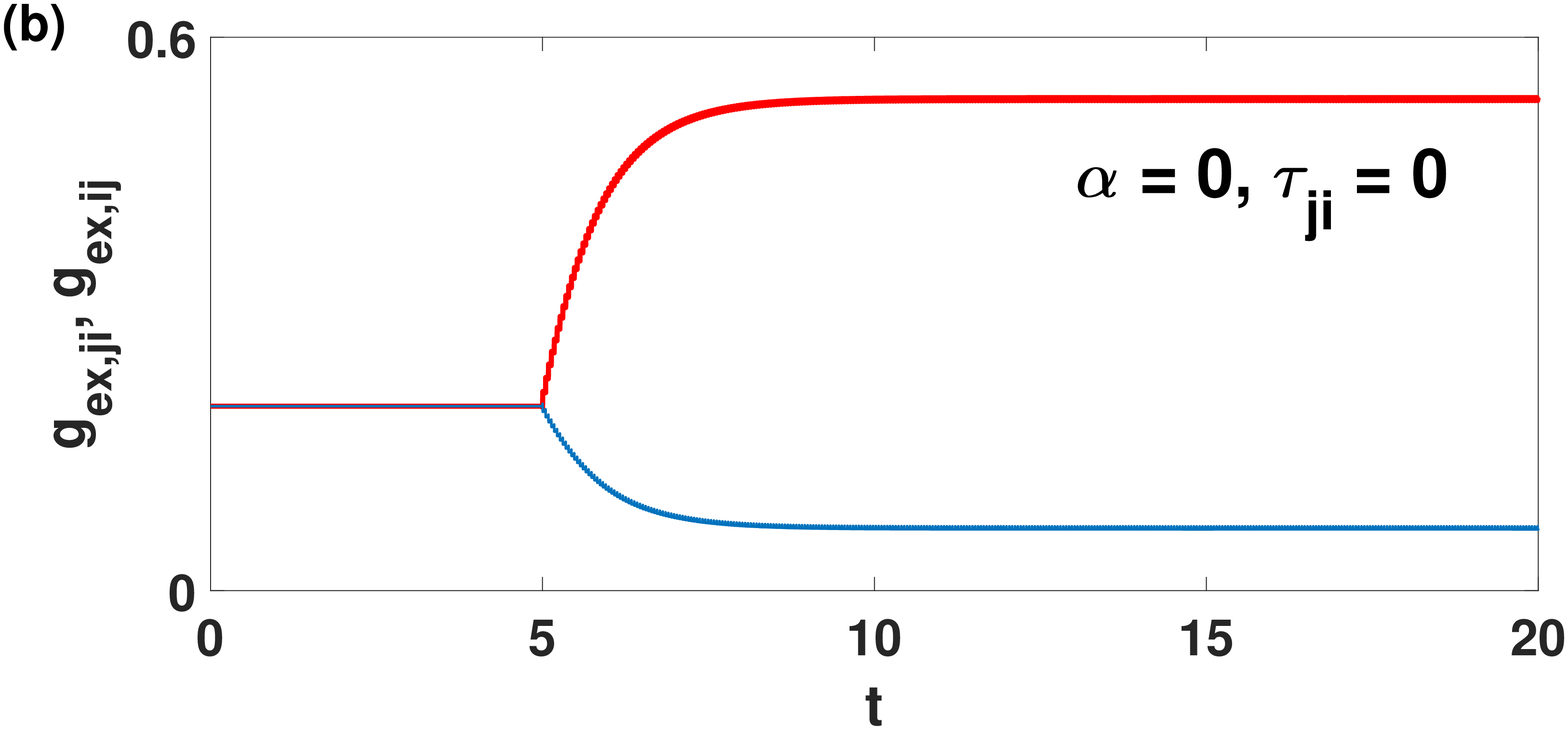}\label{fig4b}}
\subfigure{\includegraphics[width=0.22\textwidth,height=0.15\textwidth]{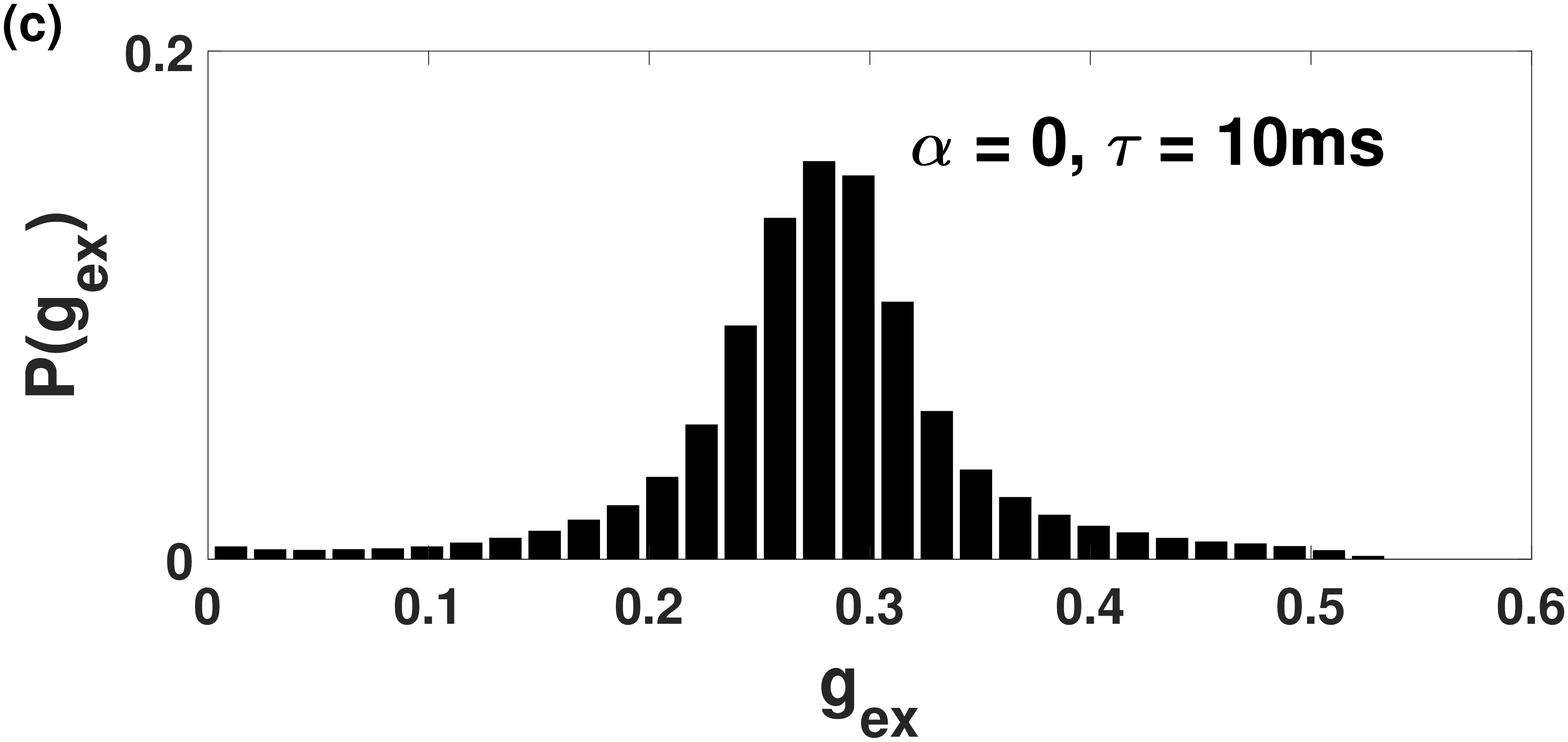}\label{fig4c}}
\subfigure{\includegraphics[width=0.22\textwidth,height=0.15\textwidth]{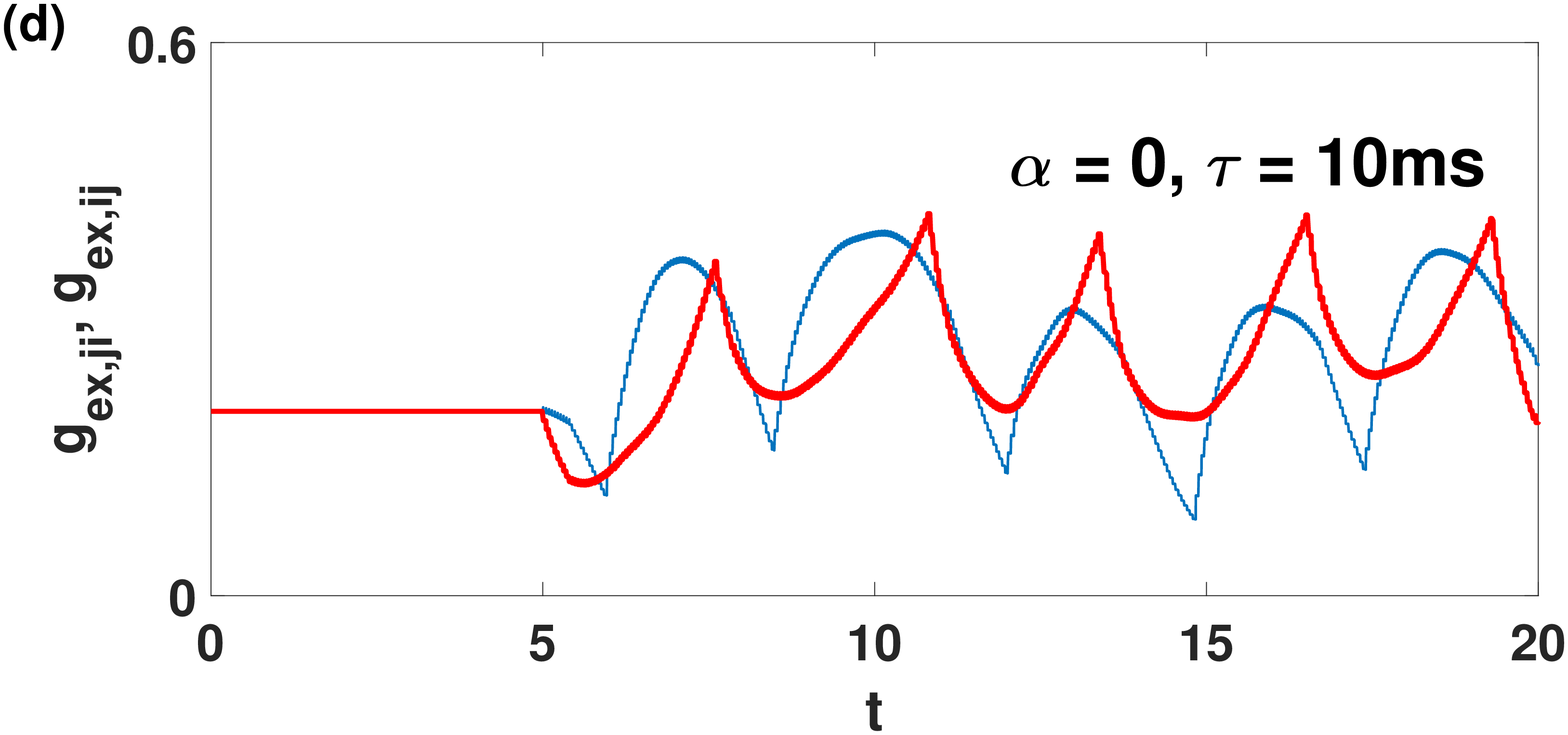}\label{fig4d}}
\subfigure{\includegraphics[width=0.22\textwidth,height=0.15\textwidth]{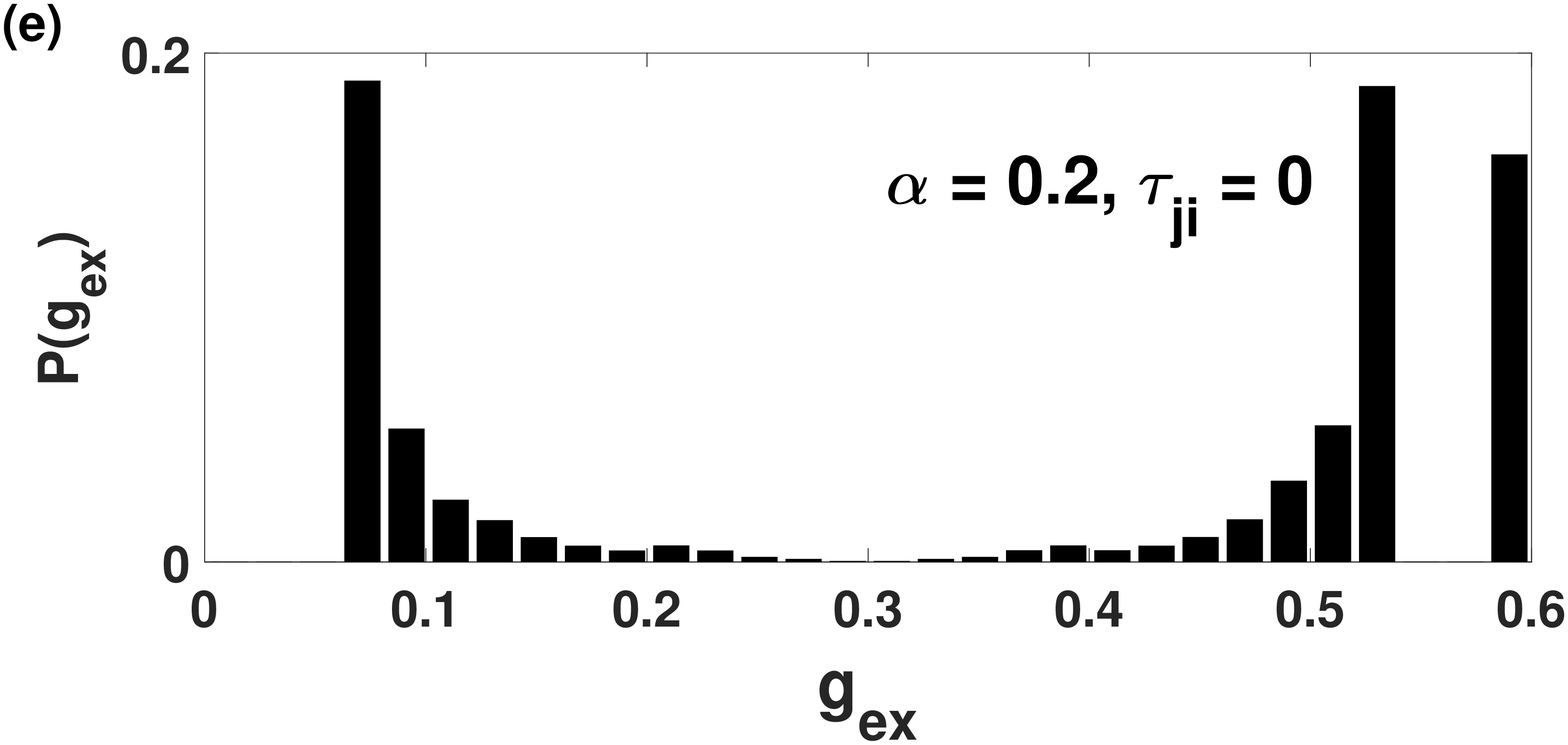}\label{fig4e}}
\subfigure{\includegraphics[width=0.22\textwidth,height=0.15\textwidth]{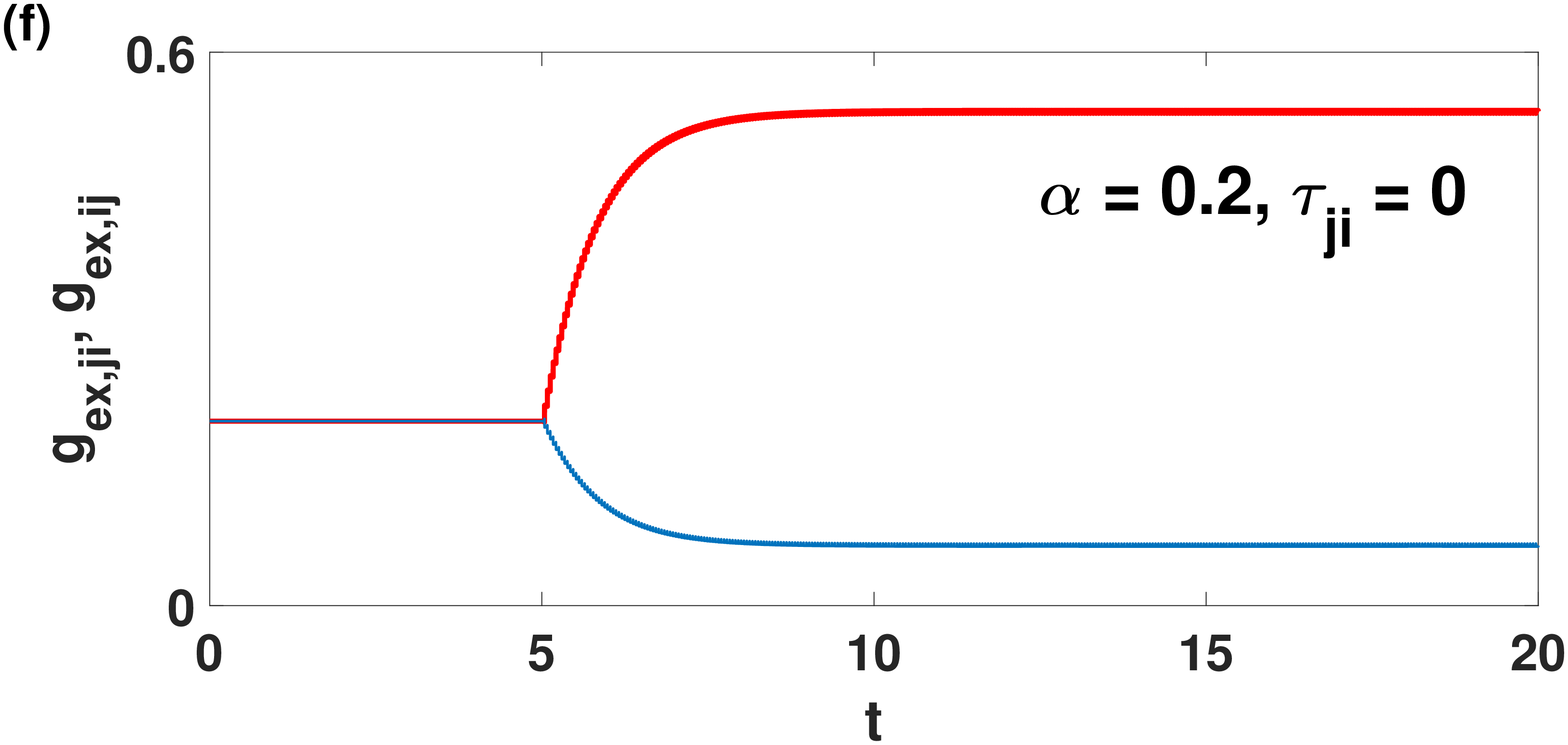}\label{fig4f}}
\subfigure{\includegraphics[width=0.22\textwidth,height=0.15\textwidth]{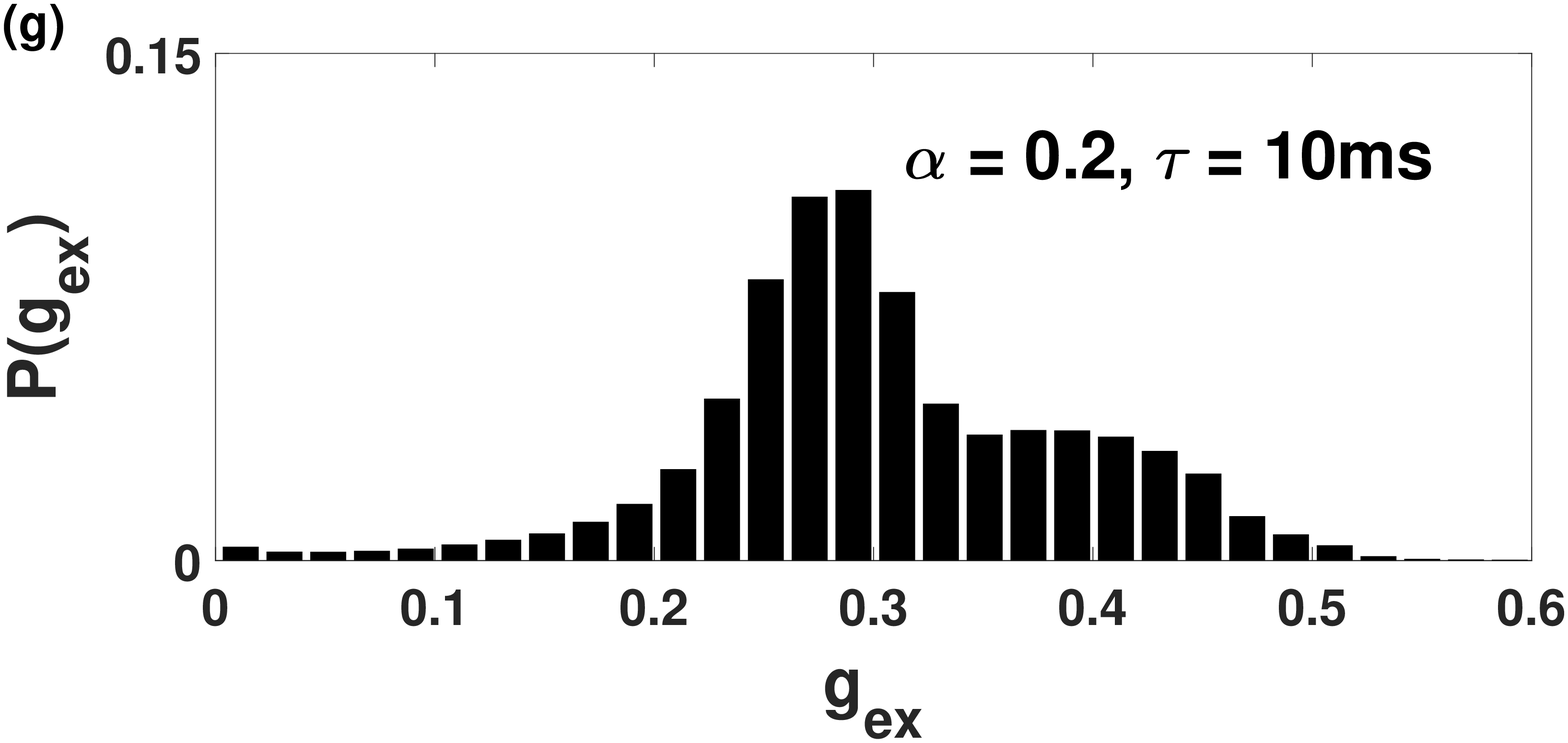}\label{fig4g}}
\subfigure{\includegraphics[width=0.22\textwidth,height=0.15\textwidth]{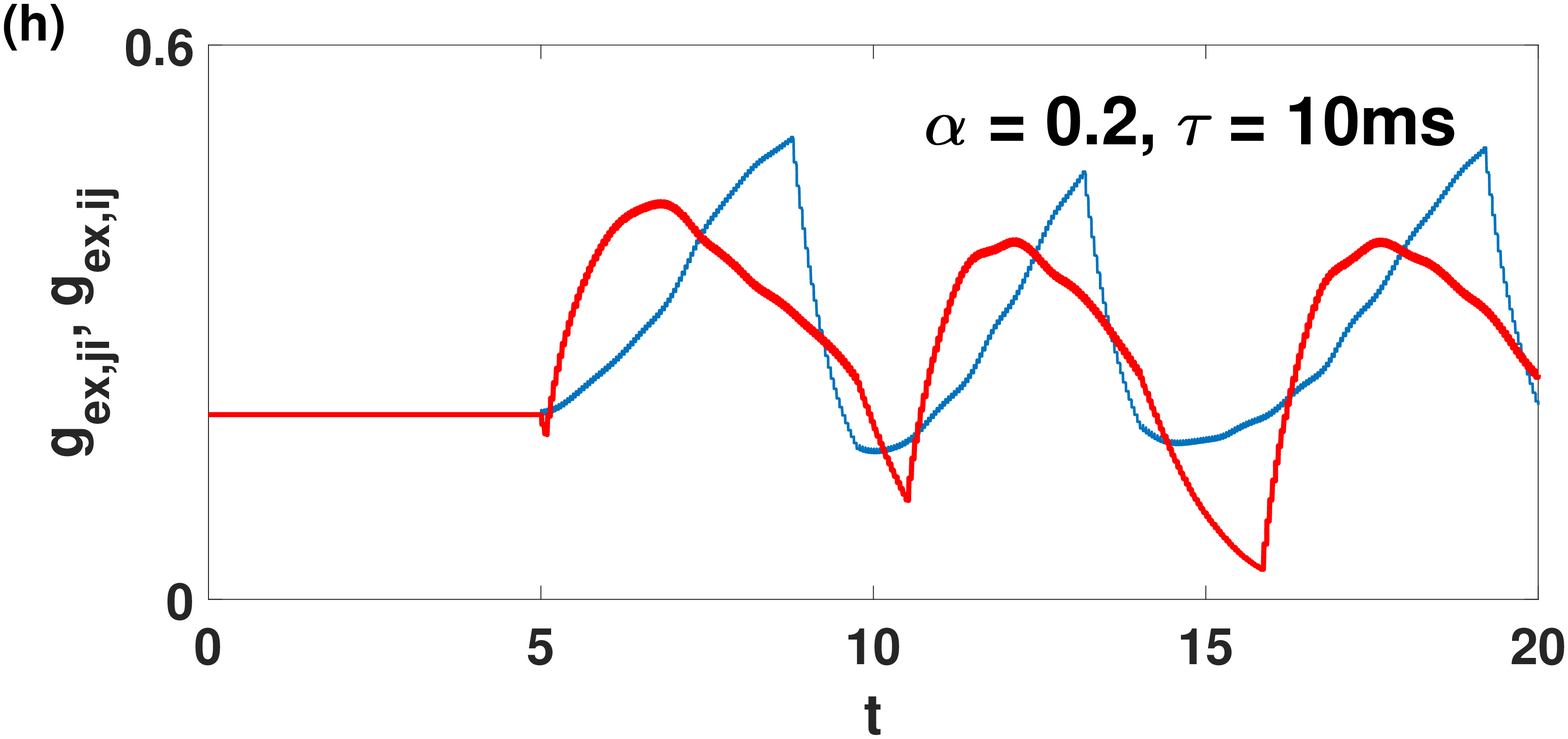}\label{fig4h}}
\end{center}
\caption{\small Distribution of the excitatory synaptic strength
at the stationary state of the systems after the influence of STDP
(left) and time evolution of a pair of reciprocal synapses
(right). The unit of time axis is in seconds.} \label{fig4}
\end{figure}

It is somewhat unexpected that similar average synaptic weights
would lead to decidedly different synchronization patterns. The
answer is in the form of the actual distributions of the weights.
In one scenario the average is the most likely value (unimodal
distribution) and in the other case is the least likely value
(bimodal distribution). The probability distribution function of
excitatory synaptic strengths $P(g_{ex})$  (in the steady state)
for each system is shown in the left column of Fig.4. Also, the
right column of this figure illustrates time evolution of strength
of a pair of reciprocal synapses.  At the absence of axonal
delays, STDP produces a bimodal distribution of synaptic strengths
(Figs.4(a) and 4(e)) which is incompatible with the experimentally
observed distributions of synaptic strength. However, addition of
time-delays to the neural network modifies this condition.
Simultaneous presence of STDP and time-delays produce a unimodal
distribution of synaptic strengths (Figs.4(c) and 4(g)) resembling
those measured in cultured and cortical networks
\cite{Turrigiano1998,Song2005}.

Emergence of these different distributions of synaptic strengths
is associated with the amount of phase synchronization in the
networks. When neurons interact without time-delay, the final
state of the system is strongly synchronized. Therefore, for each
pair of symmetric links, STDP depresses the link in one direction
and potentiates the link in the opposite direction as in Figs.4(b)
and 4(f). Thus all symmetric connection would be broken into
unidirectional connections after a while in this case. This leads
to a bimodal distribution of synaptic strengths whether the
network consists of purely excitatory neurons or a mixture of
excitatory and inhibitory neurons. With the inclusion of
time-delay in the system the level of order declines as it also
causes to preserve symmetric connections between each pair of
neurons \cite{Madadi}. Although the strength of synapses
fluctuates over time (Figs.4(d) and 4(h)), both links in opposite
directions remain active. This leads to a unimodal distribution of
synaptic strengths.

\subsection{Indicators of criticality}
\begin{figure}
\begin{center}
\subfigure{\includegraphics[width=0.22\textwidth,height=0.15\textwidth]{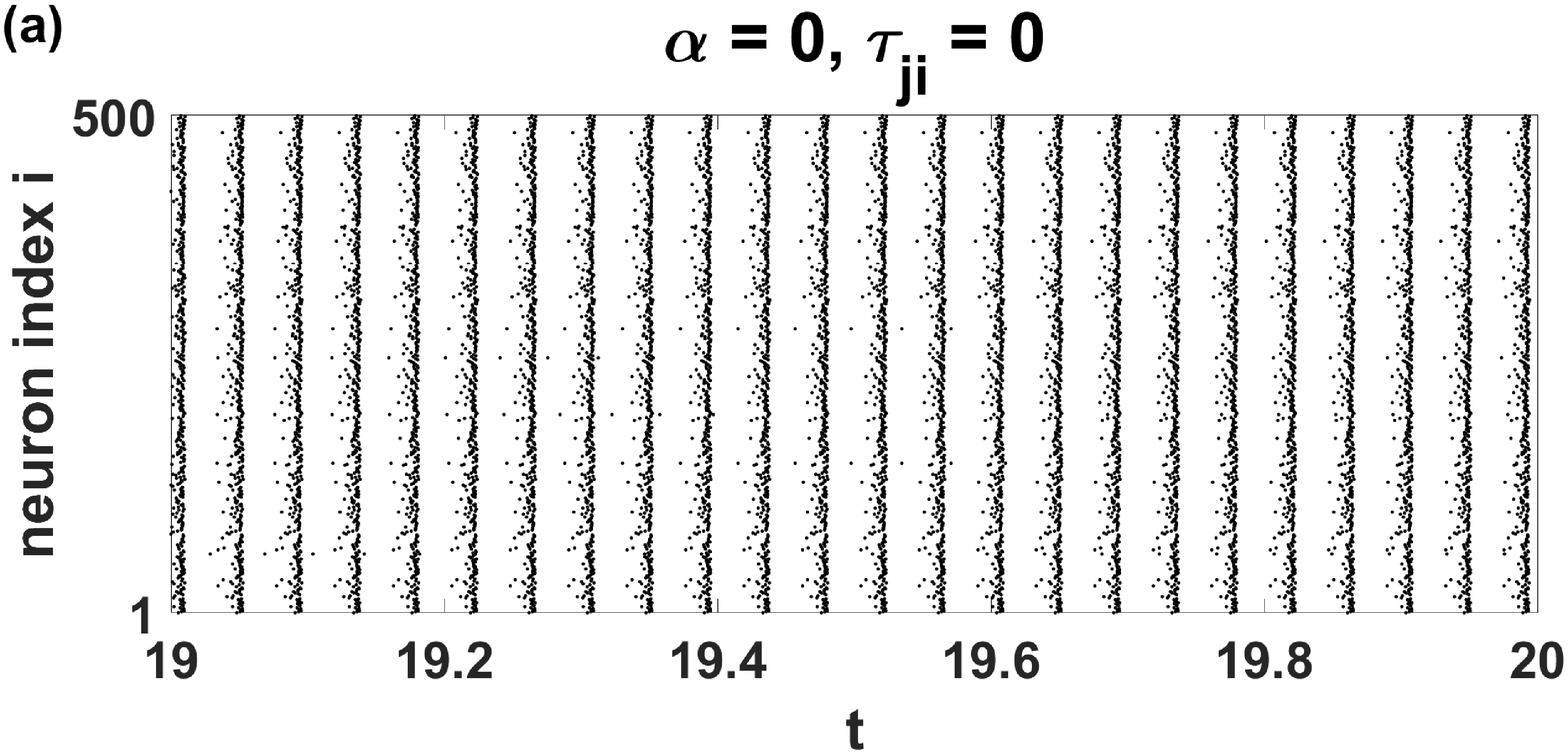}\label{fig5a}}
\subfigure{\includegraphics[width=0.22\textwidth,height=0.15\textwidth]{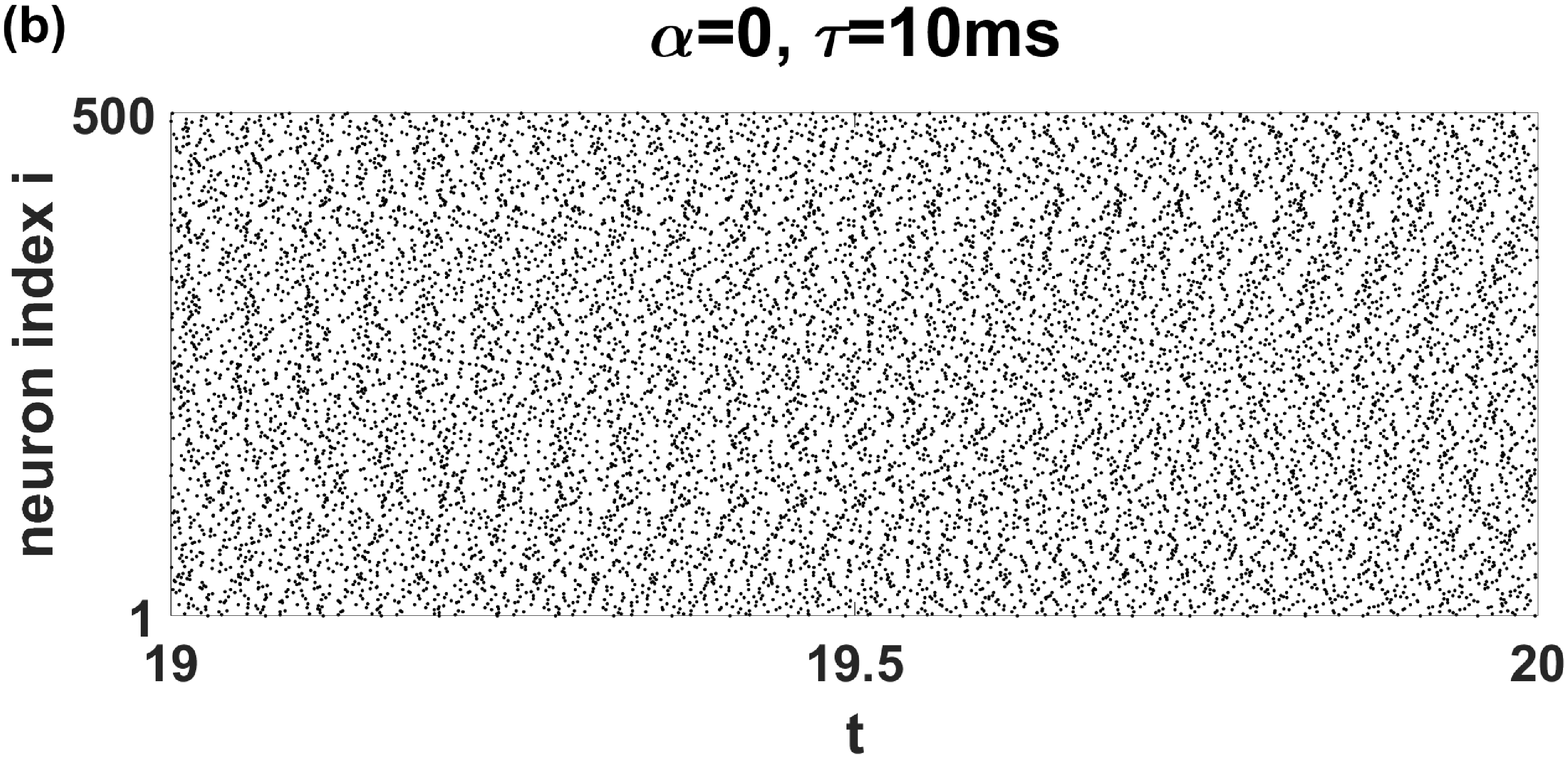}\label{fig5b}}
\subfigure{\includegraphics[width=0.22\textwidth,height=0.15\textwidth]{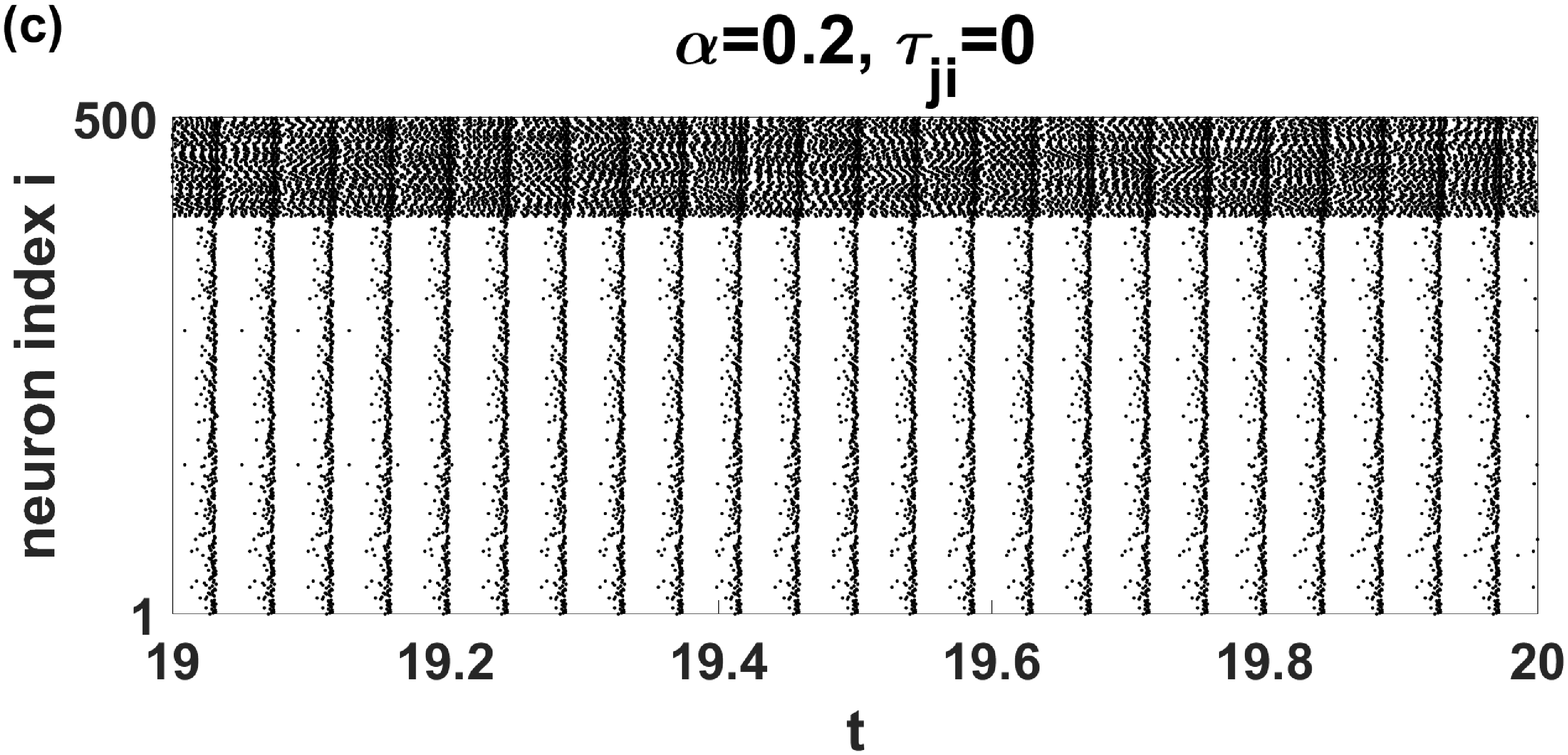}\label{fig5c}}
\subfigure{\includegraphics[width=0.22\textwidth,height=0.15\textwidth]{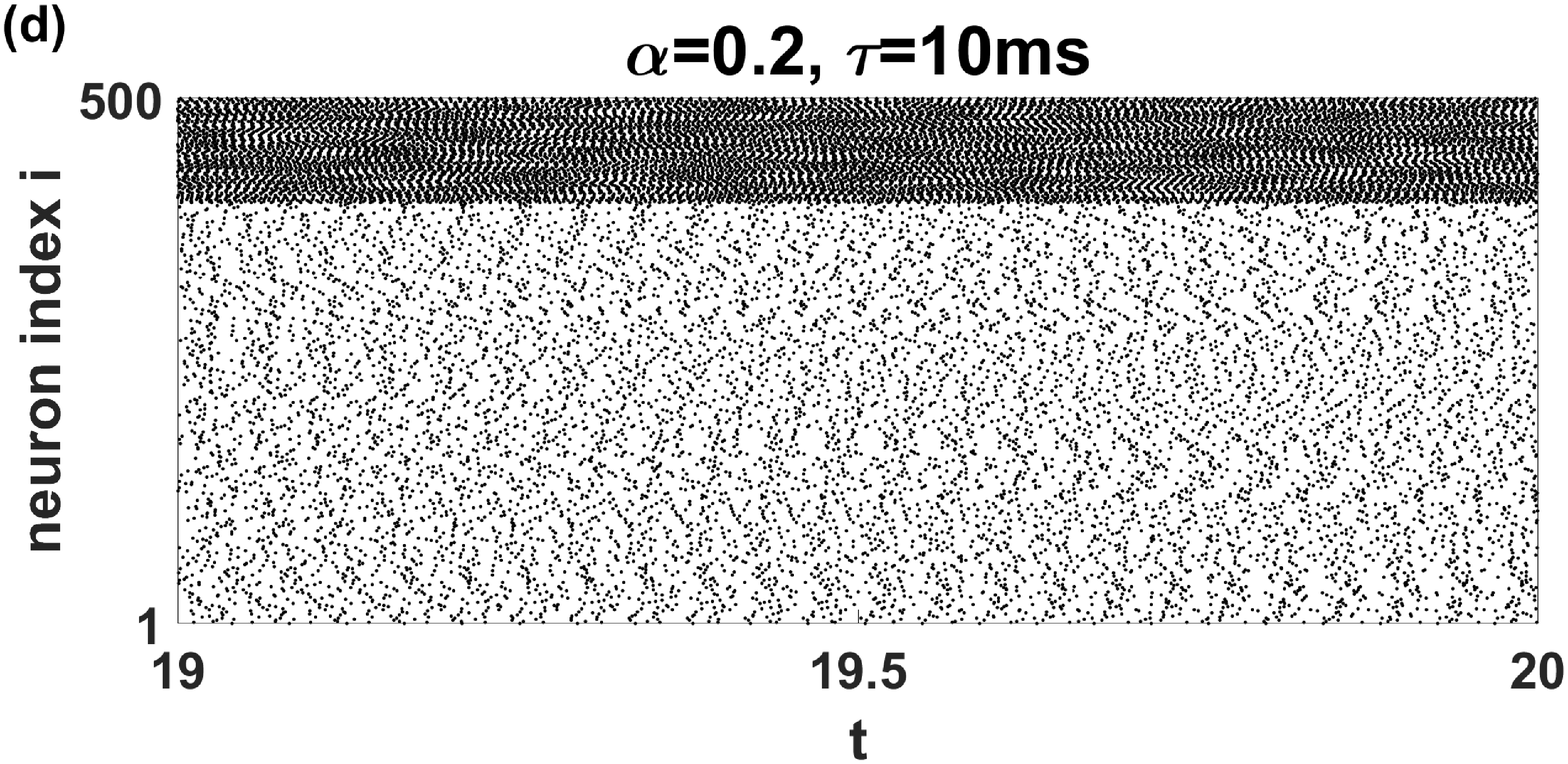}\label{fig5d}}
\end{center}
\caption{\small Raster plots of the neural networks with different
values of $\tau$ and $\alpha$ at the stationary states after
influence of STDP. The unit of time axis is in seconds.}
\label{fig5}
\end{figure}

So far we have seen that STDP along with reasonable time delay
(and inhibition) will lead the system on the edge of
synchronization. However, being on the edge of synchronization
could be caused by vastly different spiking patterns
\cite{KM2018}. More importantly for the purpose of the present
study, we would like to know whether such a state of minimal
synchronization has any experimentally relevant indications of
criticality. In this subsection we will address such issues.

Raster plots of neural networks with different values of $\alpha$
and $\tau$ (in the steady state) are displayed in Fig.5. When
time-delay is ignored, neuronal spikes are highly ordered
(Figs.5(a) and 5(c)). This is not the real state of a healthy
nervous system. However, addition of axonal conduction delay
modifies the amount of global order in the networks. Simultaneous
effect of STDP and a suitable axonal conduction delays decrease
global coherence in neural oscillations (See Figs.5(b) and 5(d)).
In Figs.5(c) and 5(d), inhibitory neurons indexed as $401-500$,
spike with a higher rate as compared to excitatory neurons
$1-400$.

\begin{figure}[!htbp]
\begin{center}
\subfigure{\includegraphics[width=0.22\textwidth,height=0.15\textwidth]{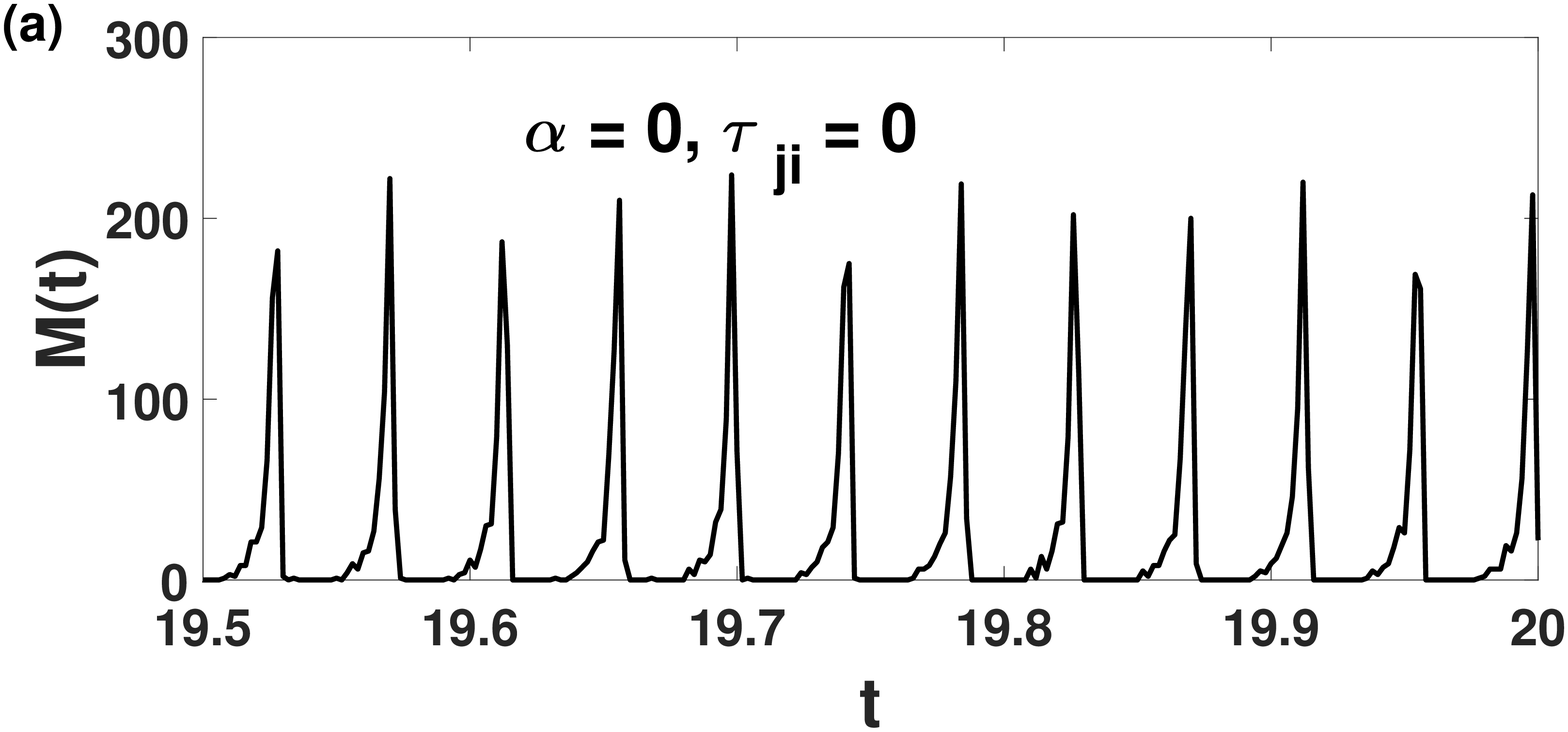}\label{fig6a}}
\subfigure{\includegraphics[width=0.22\textwidth,height=0.15\textwidth]{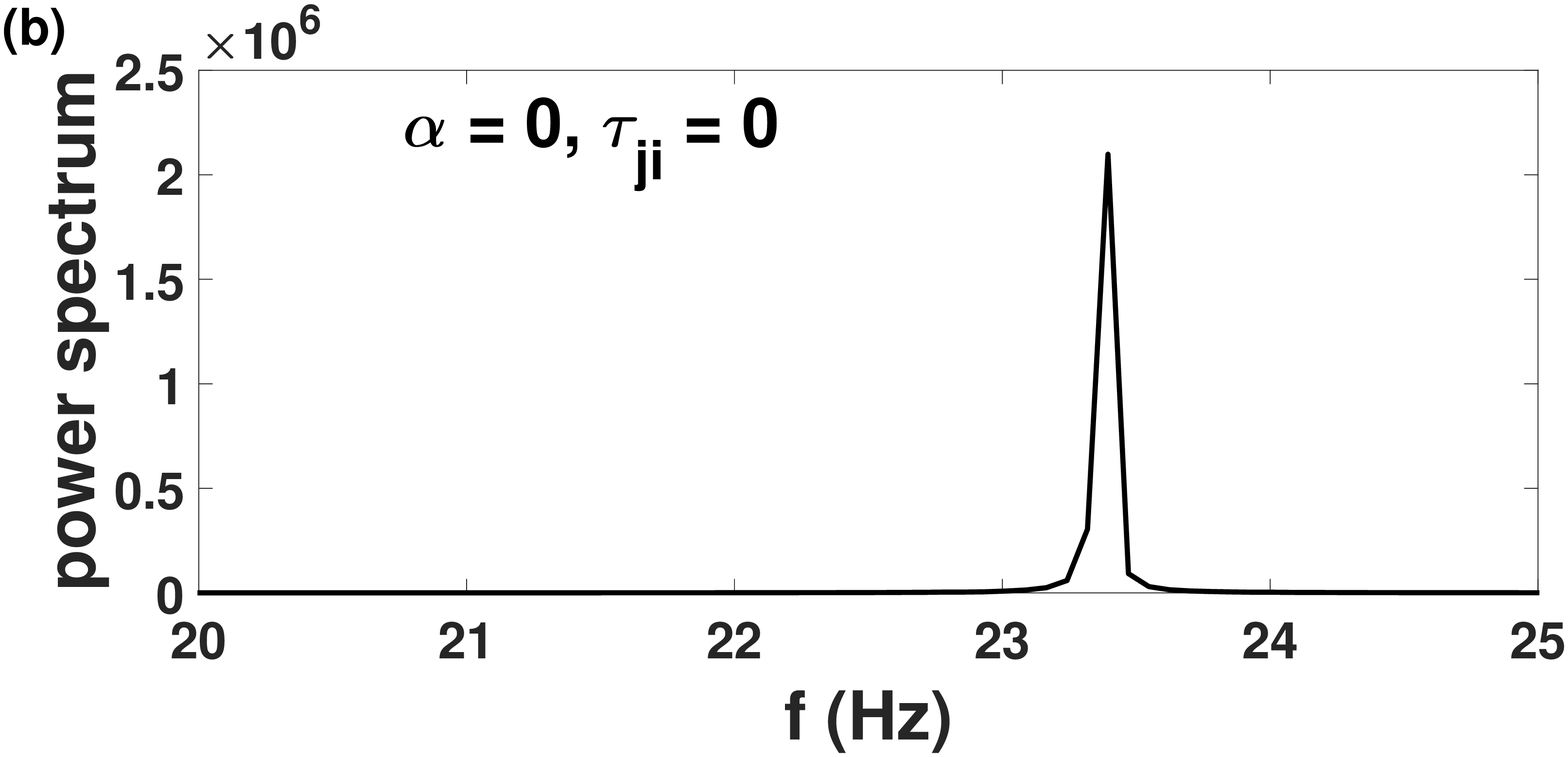}\label{fig6b}}
\subfigure{\includegraphics[width=0.22\textwidth,height=0.15\textwidth]{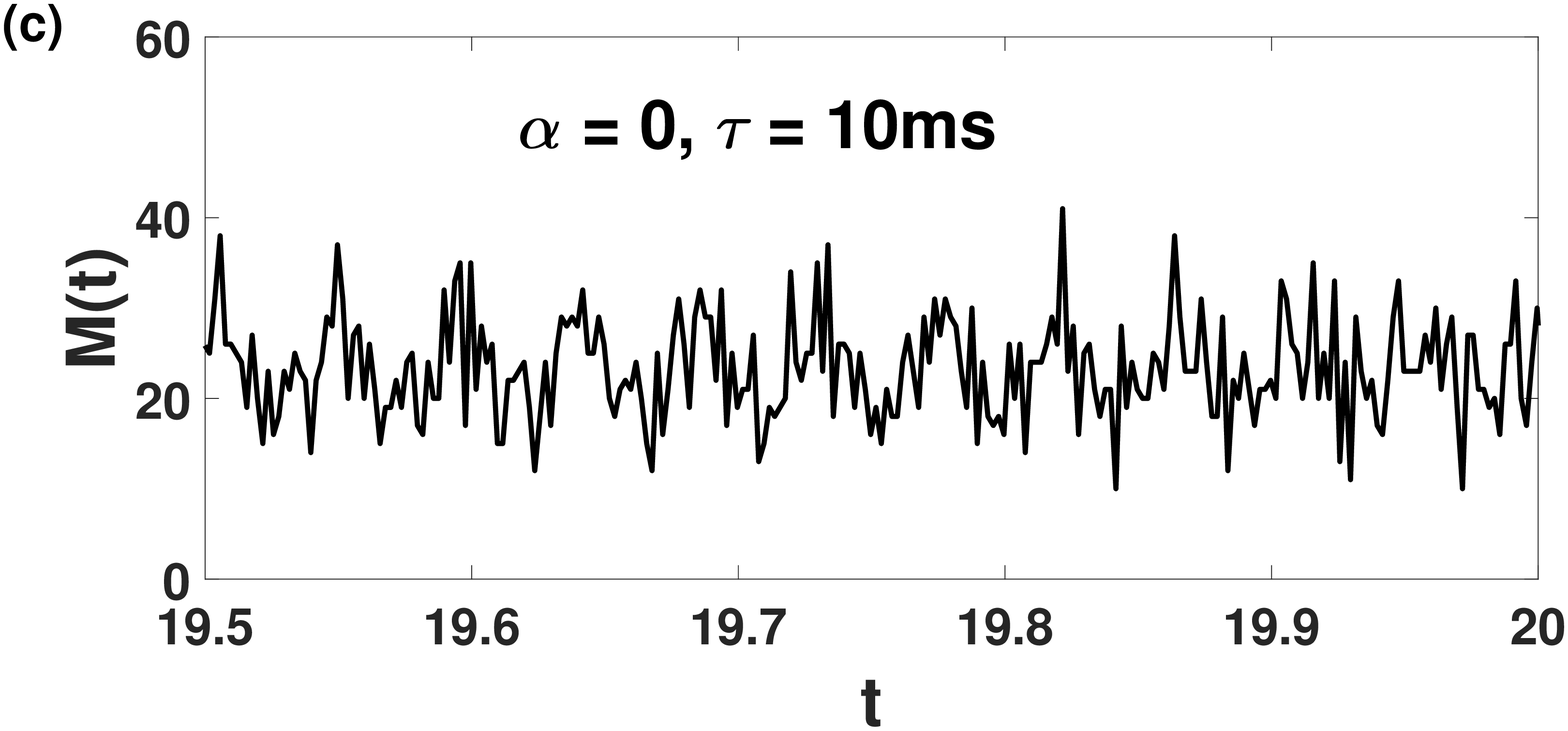}\label{fig6c}}
\subfigure{\includegraphics[width=0.22\textwidth,height=0.15\textwidth]{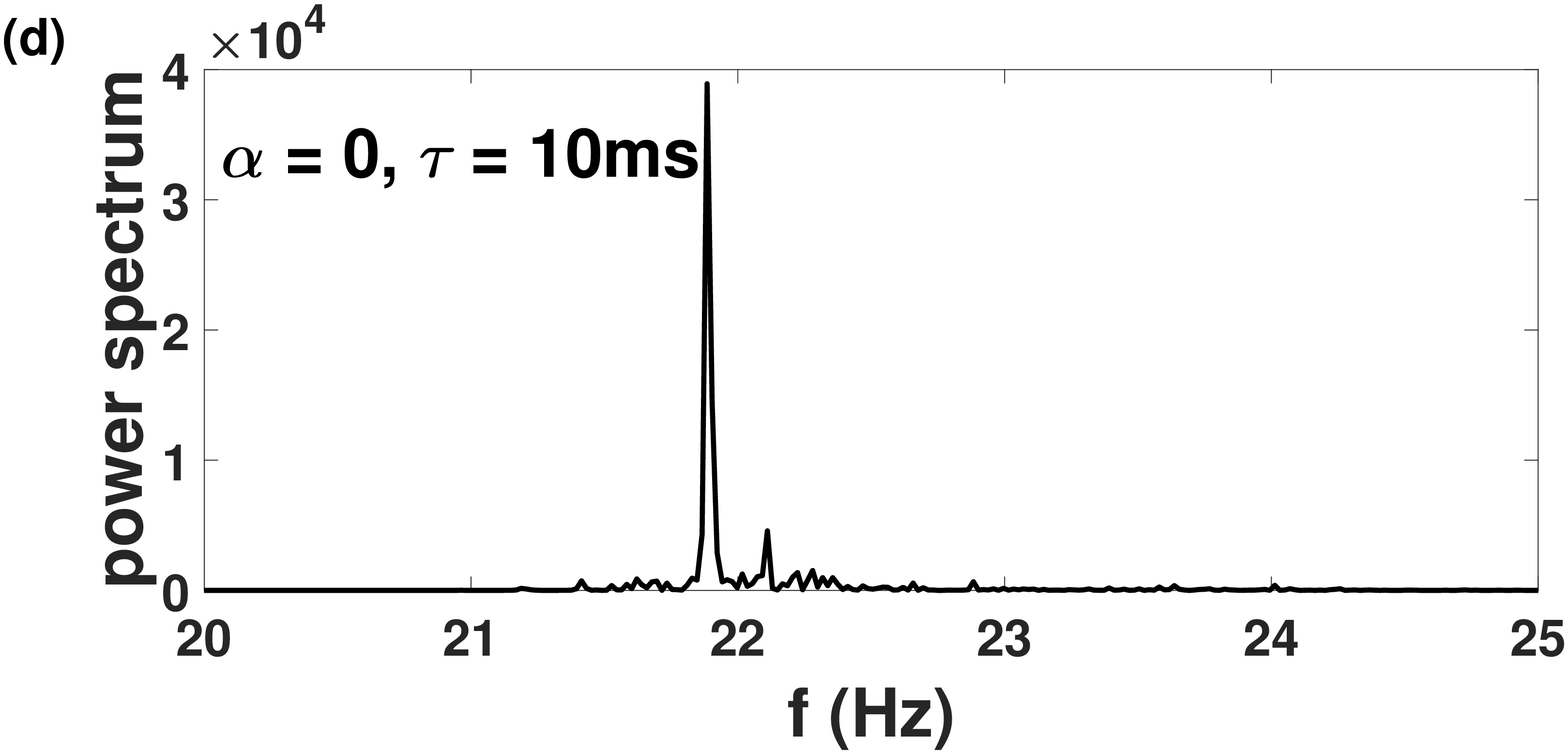}\label{fig6d}}
\subfigure{\includegraphics[width=0.22\textwidth,height=0.15\textwidth]{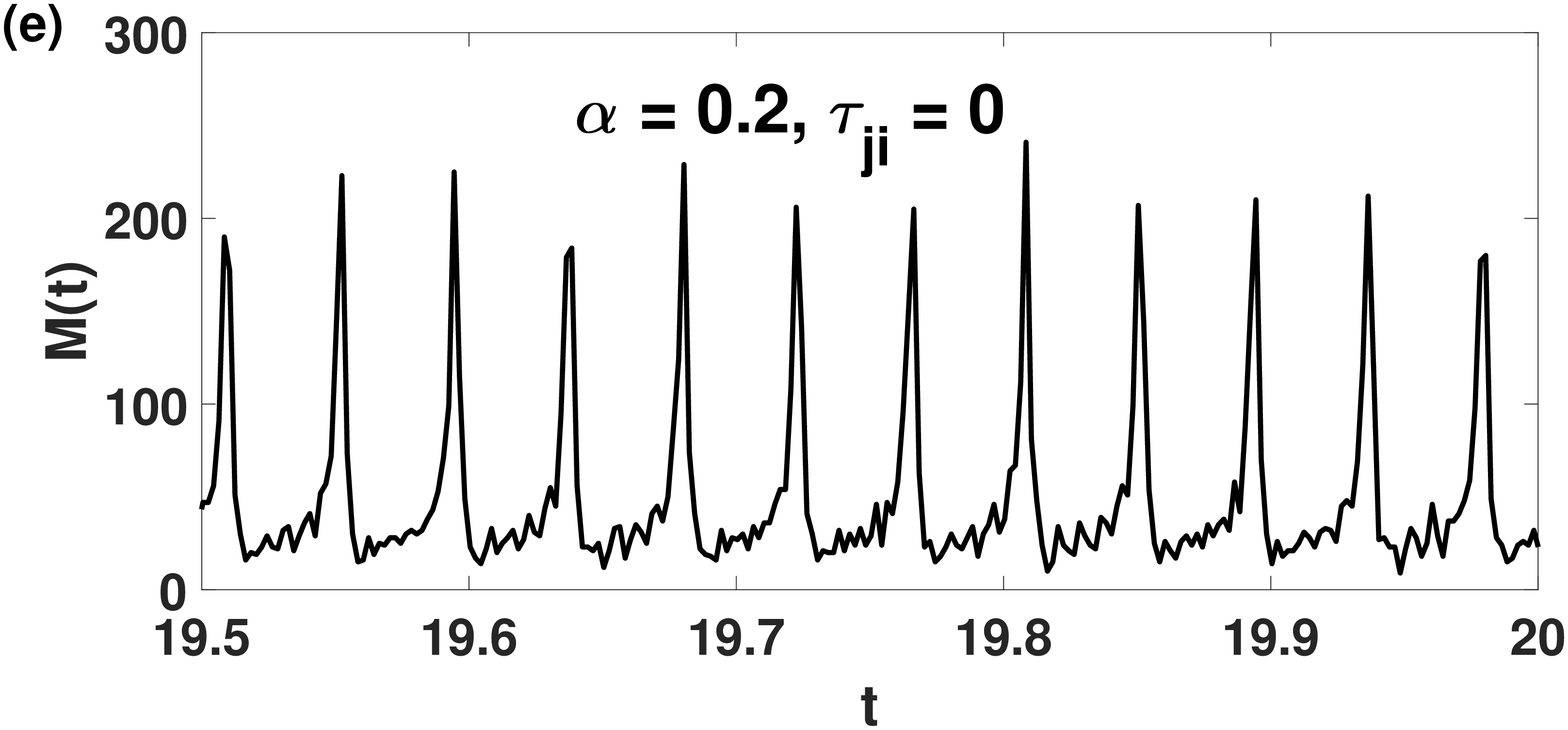}\label{fig6e}}
\subfigure{\includegraphics[width=0.22\textwidth,height=0.15\textwidth]{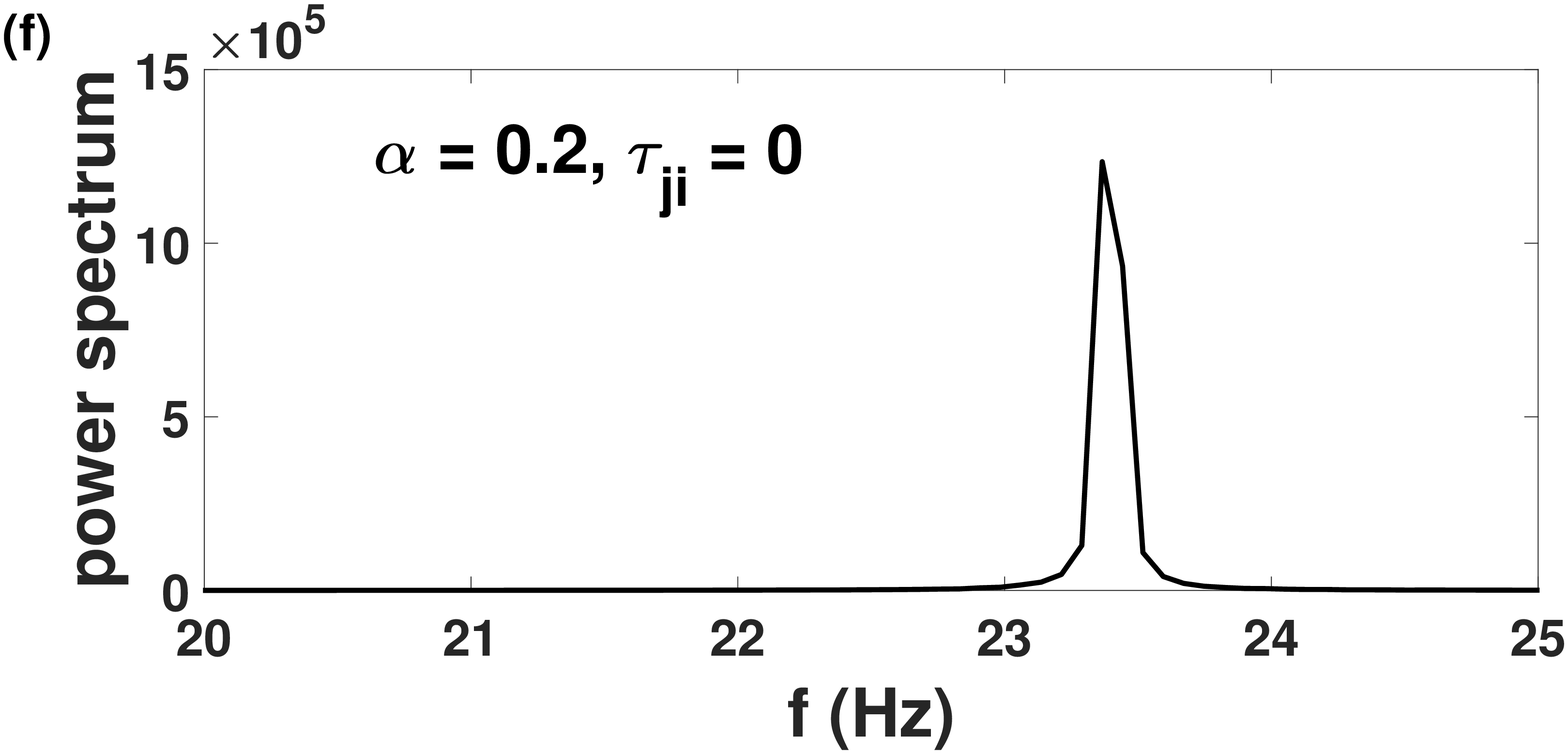}\label{fig6f}}
\subfigure{\includegraphics[width=0.22\textwidth,height=0.15\textwidth]{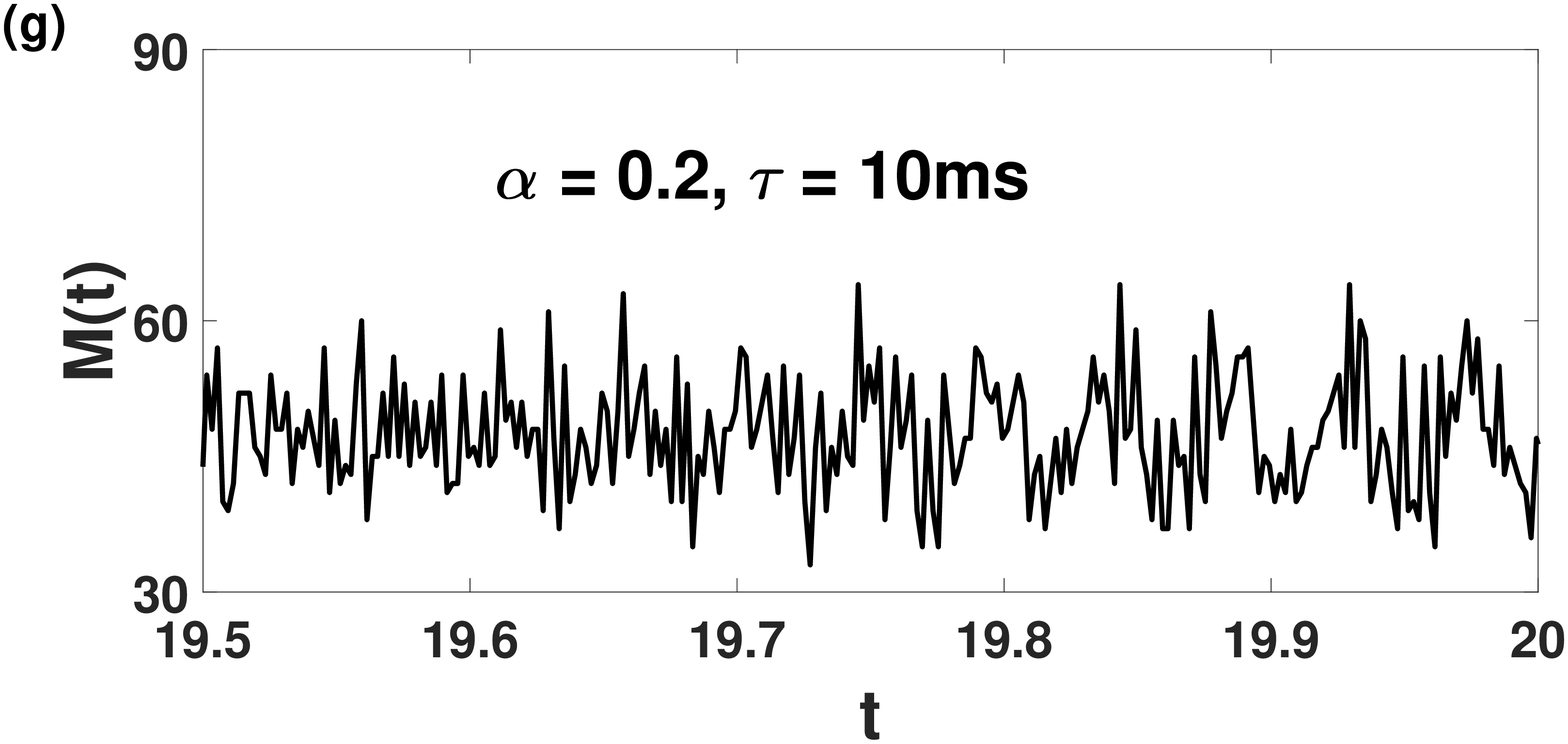}\label{fig6g}}
\subfigure{\includegraphics[width=0.22\textwidth,height=0.15\textwidth]{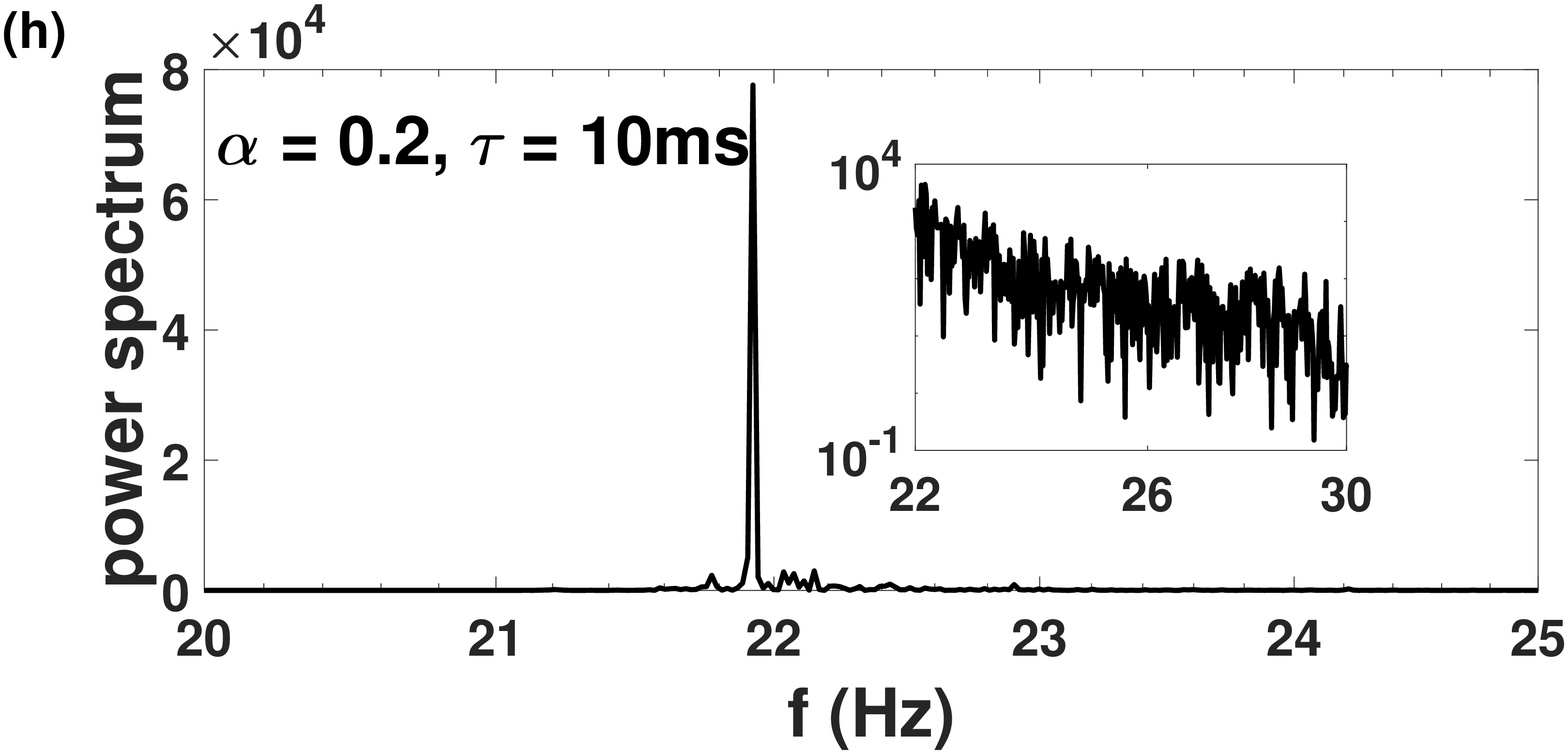}\label{fig6h}}
\end{center}
\caption{\small Timeseries of network activity $M(t)$ in the
stationary state after the influence of STDP (left) and their
power spectrum (right). The unit of time axis is in seconds. The
inset of (h) shows the same on a log-log scale.} \label{fig6}
\end{figure}

The amount of order parameter $S^*$ and the raster plots are
reasonable evidences indicating the system with $\tau=10
\text{ms}$ organizes to the edge of synchronization transition
with minimal value of $S^*$. We now present experimentally
relevant results which indicate that such a system is in a
critical state. We first consider the network activity timeseries
$M(t)$ which is defined as the number of neuronal spikes at time
$t$, as well as its power spectrum. These plots are illustrated in
Fig.6. The network activity oscillates regularly in systems
without time-delay for which phase synchronization is strong
(Figs.6(a) and 6(e)). Therefore the power spectrum of these
systems exhibit a sharp peak at $f\simeq23.5\text{Hz}$ (Figs.6(b)
and 6(f)). While neurons are delay-coupled the oscillations of
$M(t)$ are irregular (Figs.6(c) and 6(g)). Despite this deceptive
irregularity, the power spectrum exhibits a large peak at
frequency $f\simeq21.5\text{Hz}$ (Figs.6(d) and 6(h)) along with a
range of other frequencies. This dominant peak reveals that
rhythmic oscillations are still robust at these neural networks.
The inset of Fig.6(h) shows a log-log plot which indicates that
the spectrum has a decaying tail in the system for which
$\alpha=0.2$ and $\tau=10\text{ms}$. Note that the amplitude of
oscillations of $M(t)$ depends on the level of phase
synchronization. The stronger the neurons are synchronized, the
larger is the amplitude of $M(t)$ oscillations, i.e. note the
scale of the power spectrum on the y-axis.

Scale-invariant statistics of neural avalanches is thought to be
the most important indicator of critical brain dynamics. Hence,
the network displays spontaneous activity of various sizes $s$,
known as neural avalanches, which exhibit scale-free distribution,
i.e. $P(s)\sim s^{-\beta}$ \cite{Plenz2014}. By monitoring the
spiking activity of our systems, we can identify outbursts of
spikes the number of which is associated with the size of
avalanches. An avalanche begins when the network activity exceeds
a threshold $M_{th}$ and ends when it turns back below that
threshold. Here, we set the threshold to be equal with the mean
value of activity in the system. $s$ is defined as the total
number of spikes during this avalanche. Criticality is supposed to
be indicated by a power-law behavior $P(s)\sim s^{-\beta}$ and a
finite-size cut-off which diverges as system size diverges
($N\rightarrow\infty$).

We consider neural networks with $\alpha=0.2$ and three different
$\tau$ values, i.e. $\tau=14\text{ms}$, $\tau=10\text{ms}$ and
$\tau=8\text{ms}$. From the synchronization point of view, Fig.
2(b), these systems would be subcritical, critical, and
supercritical. Each network is also simulated with different
network sizes $N$. For any given set of parameters the network is
simulated for a considerably long time, producing a large number
of avalanches. Probability distribution functions of avalanche
sizes for such networks is illustrated in the left column of
Fig.7. For neural networks with $\tau=14\text{ms}$, $P(s)$ decays
with a characteristic scale which is an indicator of subcritical
behavior (Fig.7(a)). Note how this scale saturates as system size
increases. For networks with $\tau=8\text{ms}$, $P(s)$ exhibits a
bump for large $s$ which is an evidence of supercritical behavior
(Fig.7(e)). Here, large avalanches are more likely to occur than
intermediate size avalanches. Interestingly, in networks with
$\tau=10\text{ms}$ $P(s)$ exhibits a power-law behavior $P(s)\sim
s^{-\beta}$ and a finite-size cut-off which increases relative to
the system size (Fig.7(c)).

Emergence of power-law behavior in a finite system does not
necessarily prove criticality of the system. To verify
criticality, we  perform a finite-size scaling of our data for
different network sizes $N$ (inset of Fig.7(c)). We observe that
indeed we obtain a good collapse for the system sizes considered
in this study. Incidentally, our finite-size scaling collapse
allows us to calculate the $\beta$ value  more reliably where we
obtain the critical exponent $\beta=1.4$ which is close to the
accepted experimental value $\beta\approx1.5$ obtained by various
studies including the original neuronal avalanche study of
\cite{Beggs2003}.

\begin{figure}[!htbp]
\begin{center}
\subfigure{\includegraphics[width=0.22\textwidth,height=0.15\textwidth]{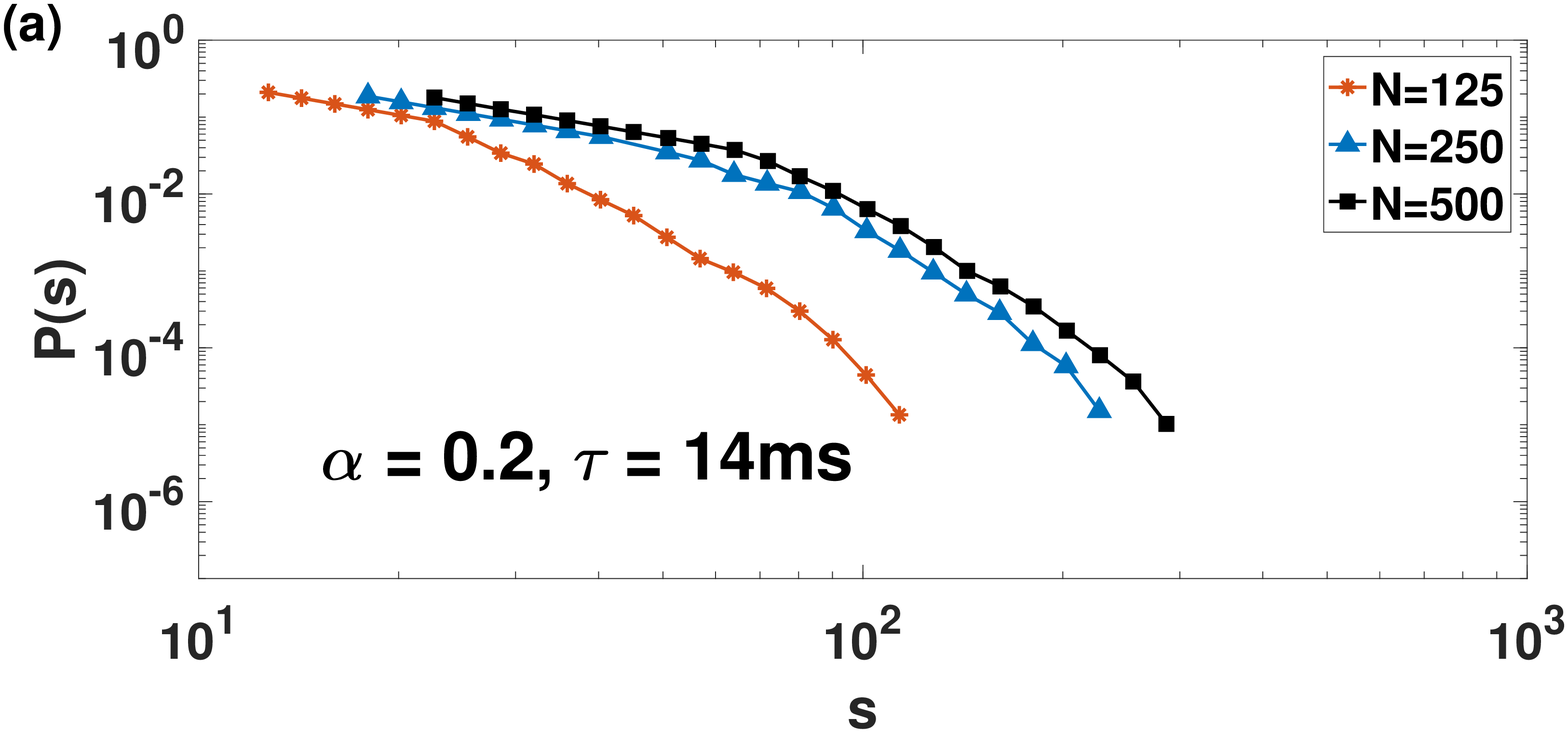}\label{fig7a}}
\subfigure{\includegraphics[width=0.22\textwidth,height=0.15\textwidth]{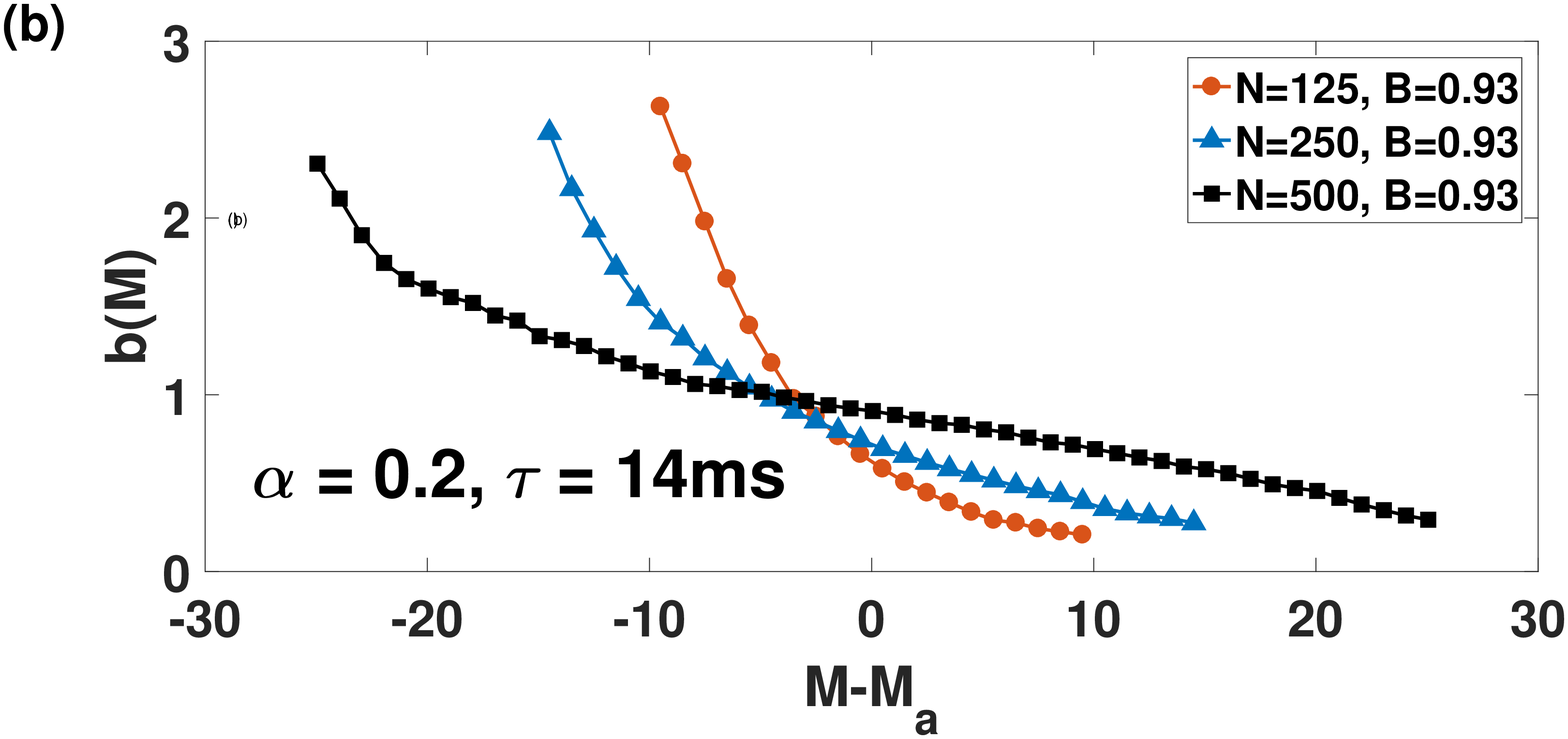}\label{fig7b}}
\subfigure{\includegraphics[width=0.22\textwidth,height=0.15\textwidth]{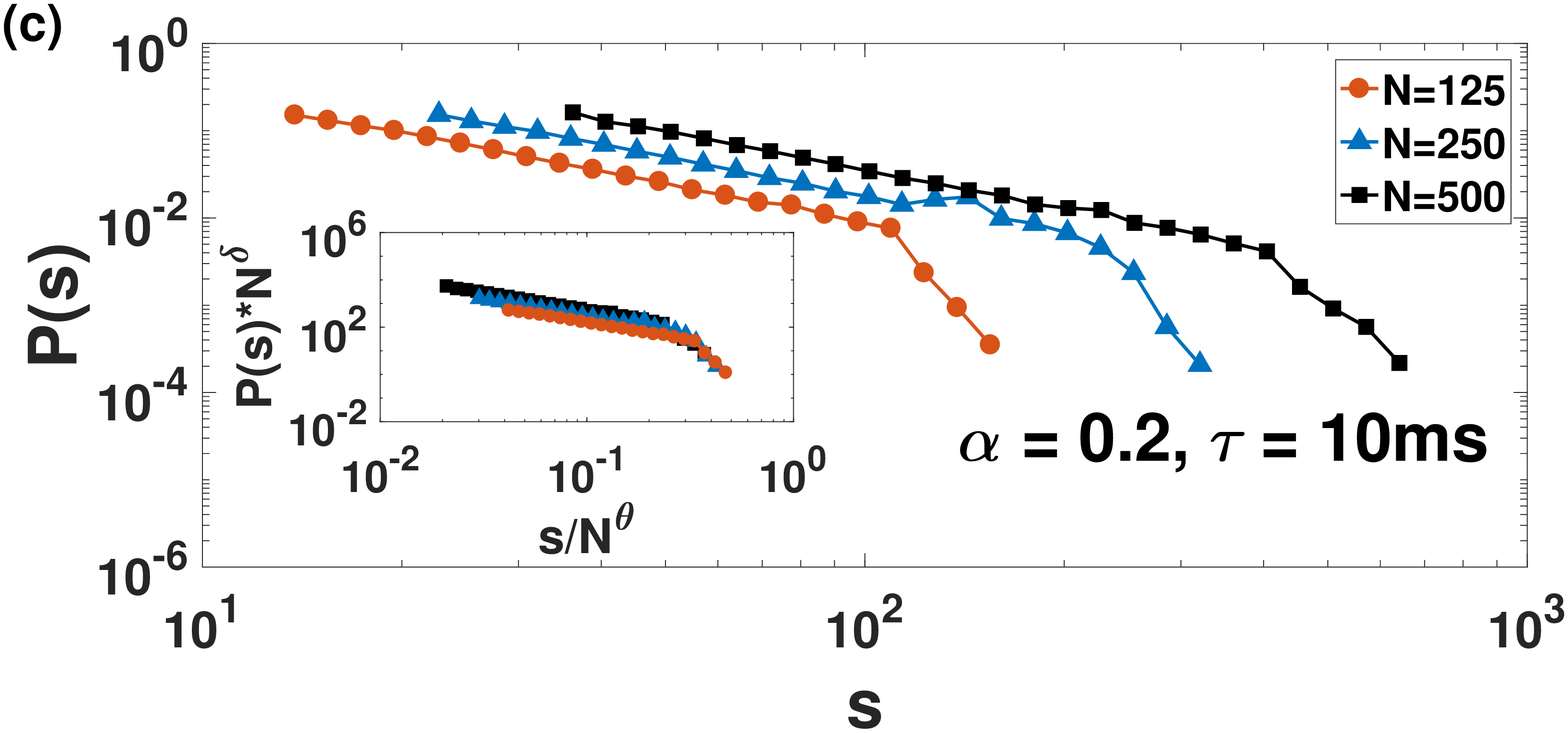}\label{fig7c}}
\subfigure{\includegraphics[width=0.22\textwidth,height=0.15\textwidth]{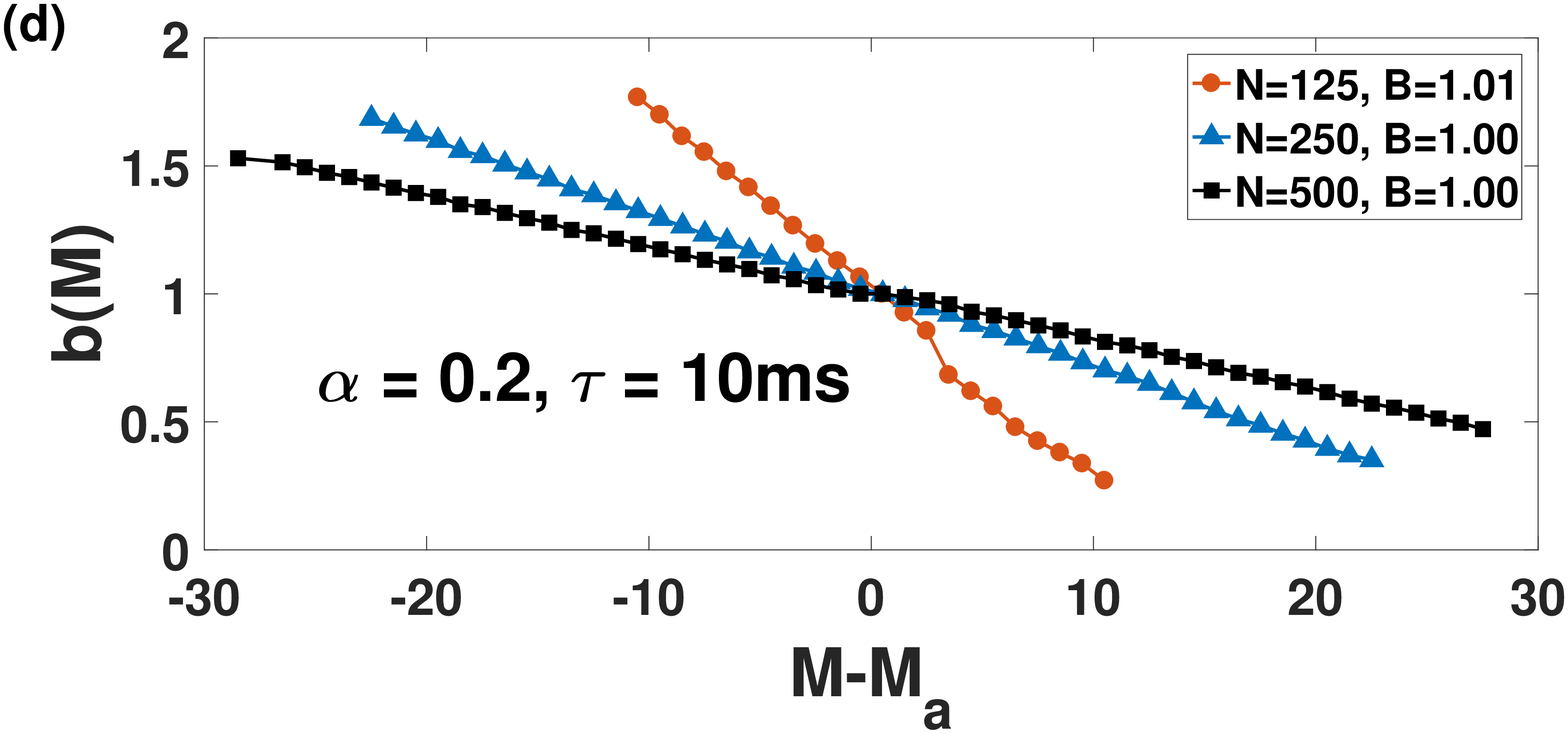}\label{fig7d}}
\subfigure{\includegraphics[width=0.22\textwidth,height=0.15\textwidth]{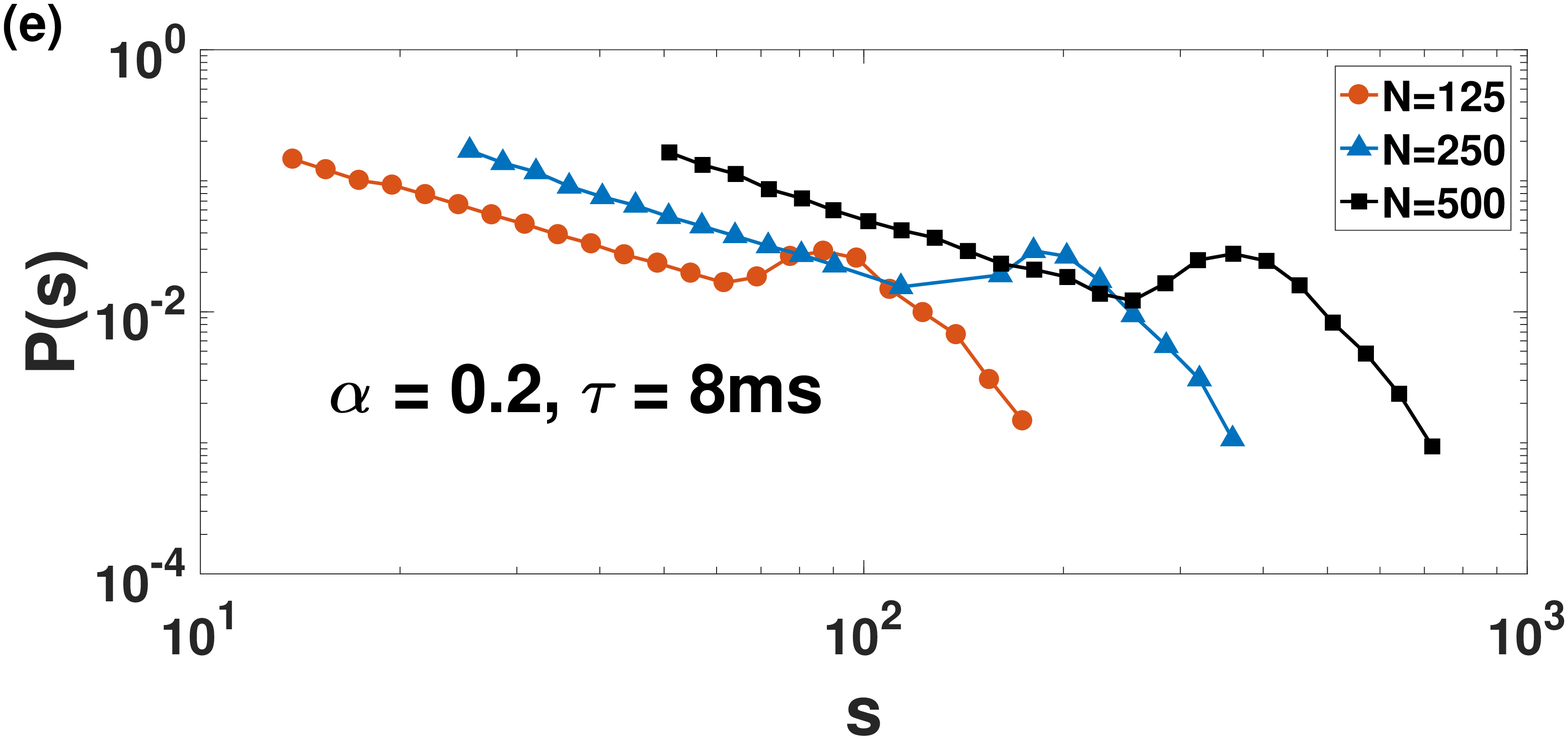}\label{fig7e}}
\subfigure{\includegraphics[width=0.22\textwidth,height=0.15\textwidth]{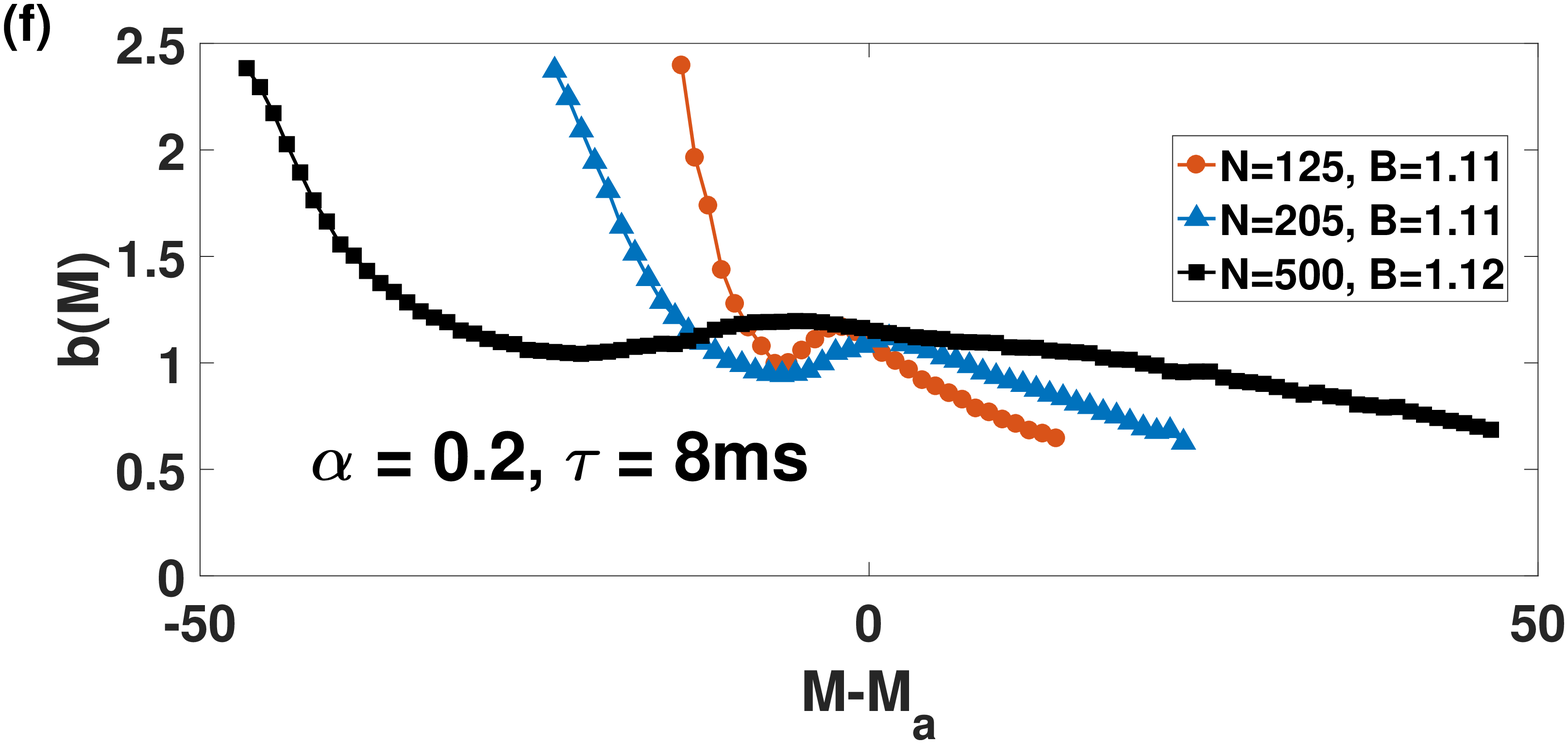}\label{fig7f}}
\end{center}
\caption{\small Distribution function of size of avalanches, and
activity-dependent branching-ratio $b(M)$ vs $M-M_a$, for various
network  sizes $N$: (a) and (b) subcritical, (c) and (d) critical,
(e) and (f) supercritical. Inset of (c) shows the
finite-size-scaling collapse of size of avalanches in the critical
system for $\theta=1.20$ and $\delta=1.68$. Thus the critical
exponent is $\beta=\frac{\delta}{\theta}=1.4$. On the right
column, the average branching-ratio $B$ for each network with size
$N$ is reported in the legend.} \label{fig7}
\end{figure}

Another important quantity to characterize critical dynamics is
\emph{activity-dependent} branching ratio \cite{Martin2010}.
Essentially, this function gives the (relative) expectation value
of the timeseries in the next time step for a given amount of
activity at the present time step. More precisely, it is defined
as, $b(M)=E{\{} \xi_M/M {\}}$. The variable $\xi_M$ is the value
of the next signal given that the present one is equal to $M$, so
$\xi_M={\{}M(t+dt) | M(t)=M{\}}$ \cite{Martin2010}. Since a
critical system is on the edge and is inherently unpredictable,
$b(M)\approx1,   \forall M$. For a finite system one expects a
similar result with the additional consideration that, with
increasing system size,  the range of activity $M$ should increase
and that $b(M)$ should asymptotically approach 1. Therefore, one
expects $b(M)<1$ to generally indicate subcritical behavior, while
$b(M)>1$ to indicate supercritical behavior. In fact, $b(M)$ has
been used to ascertain criticality in a wide range of systems
including sandpile models of SOC or solar flares \cite{Martin2010}
as well as neural networks \cite{MMV2017,Moosavi2014}.

We obtain the activity-dependent branching-ratio $b(M)$ using
timeseries $M(t)$. The right column of Fig.7 displays $b(M)$ plots
for each one of subcritical, critical and supercritical systems
for different system sizes $N$ (Figs.7(b), 7(d) and 7(f)). Note
that the plots are centered around their respective average
activity $M_a$. Only in the critical case (Fig.7(d)) do we observe
$B(M_a)=1$.  However, more importantly, we see $b(M)$ increases
its range and decreases its slope (towards zero) with increasing
system size, consistent with critical dynamics of the network. In
the two other cases, no such behavior is observed. For a more
common branching ratio, one calculates the average value of
$B(M)$, i.e.
$B=1/({M_{max}-M_{min}})\int_{M_{min}}^{M_{max}}b(M)dM$. We find
$B\simeq 1$ ($\tau = 10 ms$), $B\simeq 0.93$ ($\tau = 14 ms$), and
$B\simeq 1.1$ ($\tau = 8 ms$) again indicating critical,
subcritical, and supercritical dynamics accordingly. The average
branching ratios are reported in the legend of the corresponding
plots in Fig.7.

We have therefore shown how the system with STDP and
physiologically relevant inhibition and axonal delays will evolve
to a unimodal distribution of synaptic weights starting from a
complete uniform network. The resulting state is a state on the
edge of synchronization transition (not an activity transition)
which nevertheless shows experimentally relevant indicators of
critical dynamics including power-law avalanches with finite-size
scaling as well as branching ratios.  We also show how such
indicators of criticality disappear as one moves away from the
edge of synchronization transition via change in average delay
times.

\section{Concluding remarks}

In this paper we showed that invoking neurophysiological
regulatory mechanisms such as temporally shifted STDP and specific
amounts of axonal conduction delays ($\tau=10\text{ms}$) in a
biologically plausible model of cortical networks put the system
in a critical state at the neighborhood of synchronization
transition point. In this state the system exhibits robust
rhythmic behavior along with power-law distributions of avalanche
sizes. Furthermore, the behavior of activity-dependent
branching-ratio confirms the criticality of system in this state
as well. However for smaller or larger values of axonal conduction
delays neural networks self-organize into supercritical or
subcritical states, respectively. While the state of the network
is off-critical, neither the statistics of sizes of avalanches nor
branching-ratio exhibit the signs of criticality.

Coexistence of rhythmic oscillations and scale-invariant
avalanches is important for development of cortical layers
\cite{Gireesh2008}. Evidence for this coexistence has been found
in experimental investigations \cite{Gireesh2008,Yang2012}. Also
in theoretical studies, this phenomenon has been reported to occur
as a result of balance between inhibition and excitation
\cite{Poil2012},  as well as in a periodically driven SOC model
\cite{Moosavi2018}. The neurophysiological mechanisms leading to
this intricate dynamics in the cortex is of  fundamental
importance in neuroscience. Here, we revealed that such intricate
dynamics emerges as a result of intrinsic regulatory mechanisms
like STDP and axonal conduction delays. More strictly, we obtained
self-regulated criticality along with coexistence of rhythmic
oscillations and scale invariant activity in a biologically
relevant model.

We began this paper by posing three open questions regarding the
critical brain hypothesis. Our results have provided interesting
answers to all three questions. (i) The critical point and
corresponding phase transition that the brain organizes itself
into is not the usual activity and/or absorbing phase transition,
but the synchronization phase transition. (ii) The self-organizing
mechanism which tunes and maintains the system around such
critical point is a standard neurophysiological regulatory
mechanism of a temporally shifted STDP. (iii) The existence of
individual neuronal oscillations which self-organize to a highly
correlated but weakly synchronized collective state is responsible
for a dominate oscillatory mode in addition to scale-free
fluctuations.

We have studied neural networks with different topologies, various
initial conditions, as well as various choices of STDP parameters
and observed that our results are generally the same upon all such
changes. We have also examined that hard-bound STDP leads to
similar results, except for the distribution function of synaptic
strengths that would be bimodal regardless of all conditions
implemented in the neural network.

\section{Acknowledgements}
Support from Shiraz University research council is kindly
acknowledged. This work has been supported in part by a grant from
the Cognitive Sciences and Technologies Council.

\bibliographystyle{apsrev}
%\bibliography{xbib}

\end{document}